# Chatbot Engagement Does Not Always Beget Metalearning: Evidence from Three Countries

KOKIL JAIDKA[1,2], INSYIRAH BINTE IMAM MUJTAHID, PENG QI[3], HARSHIT ANEJA, SUBHAYAN MUKERJEE[2], WYNNE HSU[1,3,4], MONG LI LEE[1,4], and TSUHAN CHEN[1,4], [1]Centre for Trusted Internet & Community (CTIC), [2]Department of Communications and New Media, [3] Institute of Data Science, [4] School of Computing, National University of Singapore (NUS), Singapore

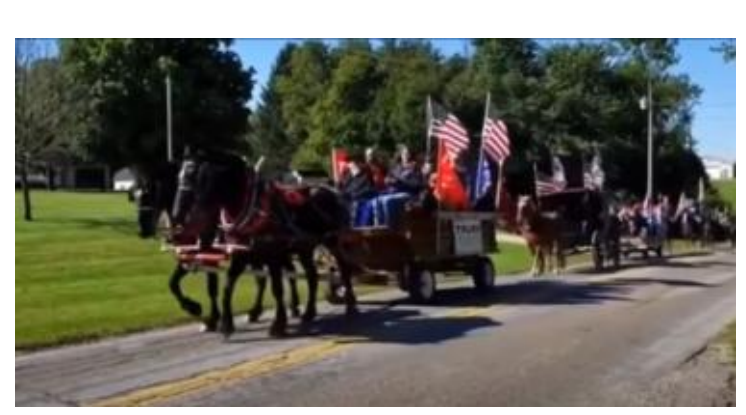

**USA**
*Original:* "Plain Parade" rallies support for the President in Amish Ohio, 2020. Source: Amish America
*Shared as:* "Over 200,000 Amish voters registered to vote for Trump in Pennsylvania in 2024."

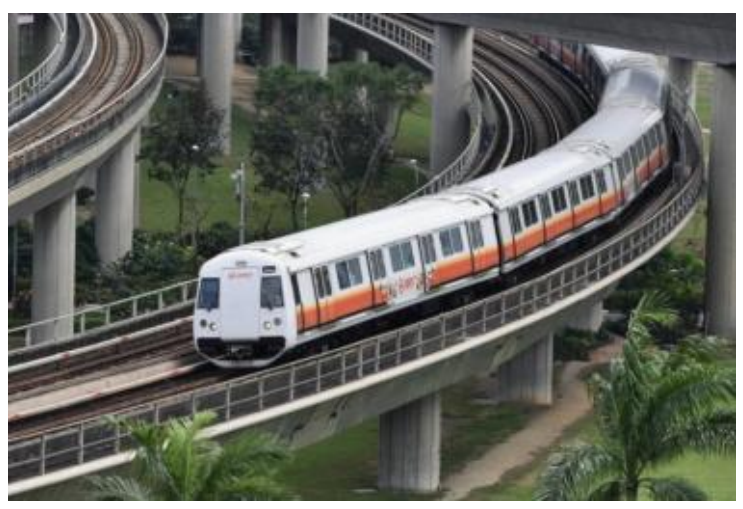

**India**
*Original:* SBS Transit and SMRT Trains apply for a fare increase. source: Straits Times Singapore
*Shared as:* "If employment hasn't increased, how did the metro reach the cities of India? Congress will say, BJP will do!"

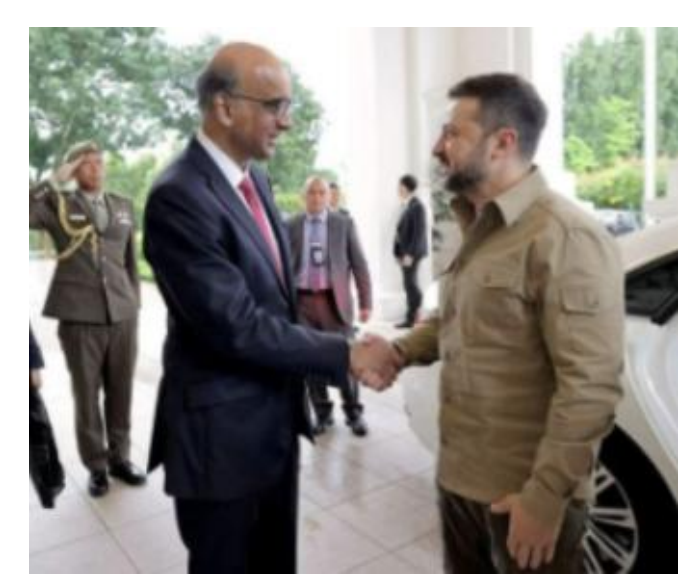

**Singapore**
*Original:* Ukrainian President Volodymyr Zelensky with President Tharman Shanmugaratnam. source: Straits Times Singapore
*Shared as:* "President Tharman meets visiting New Zealand Prime Minister Larry Laxson at the Istana."

Fig. 1. Out-of-context (OOC) misinformation items from the study, one per country. Each photograph is authentic and unedited. The misinformation occurs through linking them with false captions (provided as the "shared as" captions).

Chatbots deliver real-time fact-checks, but whether a chatbot correction leaves anything behind once the chatbot is gone – metalearning, distinct from correcting misbeliefs – is untested. We report a preregistered, three-country randomized experiment (USA, India, Singapore; $N \approx 2{,}200$) on out-of-context image misinformation, manipulating a correction's *channel affordances* (synchronicity, bandwidth) across four conditions: Control, Links-only, Static explanation, and a Socratic Chatbot built on a validated out-of-context detector, with an unaided retest one week later. The Chatbot produced the largest immediate discernment gain ($d = 0.097$, $p = .023$). All three interventions reduced sharing of false claims ($d \approx -0.12$, $p < .01$). One week later, no advantage persisted: the Chatbot arm declined relative to Control, most sharply in India and Singapore, and in India on claims it never discussed. Decay tracked affordance level and did not vary by country. Engagement mechanisms, we argue, do not substitute for slow AI literacy.

Authors' Contact Information: Kokil Jaidka[1,2], jaidka@nus.edu.sg; Insyirah Binte Imam Mujtahid, insyirah@nus.edu.sg; Peng Qi[3], pengqi.qp@gmail.com; Harshit Aneja, contactaneja@gmail.com; Subhayan Mukerjee[2], mukerjee@nus.edu.sg; Wynne Hsu[1,3,4], whsu@comp.nus.edu.sg; Mong Li Lee[1,4], leeml@comp.nus.edu.sg; Tsuhan Chen[1,4], tsuhan@nus.edu.sg, [1] Centre for Trusted Internet & Community (CTIC), and [2] Department of Communications and New Media, and [3] Institute of Data Science, and [4] School of Computing, National University of Singapore (NUS), Singapore, Singapore.

## 1 Introduction

Misinformation is a pervasive societal problem, and one of the easiest and most effective ways to spread it does not require fabricating anything at all: pairing an authentic, unaltered image with a caption that misrepresents what it depicts, known as out-of-context (OOC) misinformation [5]. Across 49 fact-checking organizations worldwide, OOC media is the single most common category of visual misinformation verified in 2026, ahead of either manipulated or AI-generated images individually [17]. During the 2023 Israel-Hamas war, for example, numerous instances of OOC misinformation circulated on social media, repurposing old images from unrelated conflicts, or even video-game footage, as evidence of current events [5]. OOC misinformation is specifically hard to catch, for machines and people alike, because there is nothing manipulated in the image itself for a detector – or a person – to find: the deception lives entirely in the mismatch between an authentic image and the claim attached to it.

As seen in Figure 1, OOC misinformation pairs a real image with a caption that misrepresents its source or context. The three countries in this study illustrate why this problem cannot be studied, or addressed, as a single global phenomenon. In the United States, 73% of respondents report being worried about distinguishing real from fake news online, matching the highest levels recorded globally [26]. Among English-speaking online news users in India, 57% report the same worry, and messaging apps – especially WhatsApp, used by 82% of this population – are a distinctively prominent misinformation concern relative to other countries [3]. In Singapore, 65.2% of adults report difficulty distinguishing authentic from false online content, above the global average, but in a media environment with substantially higher institutional trust and a more centralized regulatory response to online falsehoods than the U.S. or India [40]. Three countries, three different baseline relationships to misinformation and to the institutions meant to correct it – yet nearly all experimental evidence on what corrects misinformation, and whether any correction lasts, comes from just one of these contexts at a time (Section 2.3).

In our study, we evaluate the efficacy of a chatbot custom-built on a fact-checking system that detects and explains OOC misinformation (Section 3). Chatbots are increasingly used to perform tasks once handled by search engines and static fact-check pages: answering questions, summarizing news, and resolving disputed claims in real time. As large language models become the default interface for information-seeking, understanding what a chatbot-mediated correction actually leaves behind is an urgent design and policy question – and answering it requires separating two things a single correction encounter can do. It can produce *object-level learning*: a user revises their belief about the specific claim they were just shown corrected. It could also, much more ambitiously, produce *metalearning*: the user internalizes the demonstrated verification procedure itself – checking an image against its caption, checking the caption against the evidence, verifying the source – well enough to imitate it independently on a claim they were never walked through, in the sense of a scaffold (here, the chatbot) being removed and the learner tested on whether the modeled procedure survives without it [13]. Prior work on fact-checking interventions rarely distinguishes these two constructs explicitly, but they have different implications: an intervention can be a complete success at the first and entirely silent on the second.

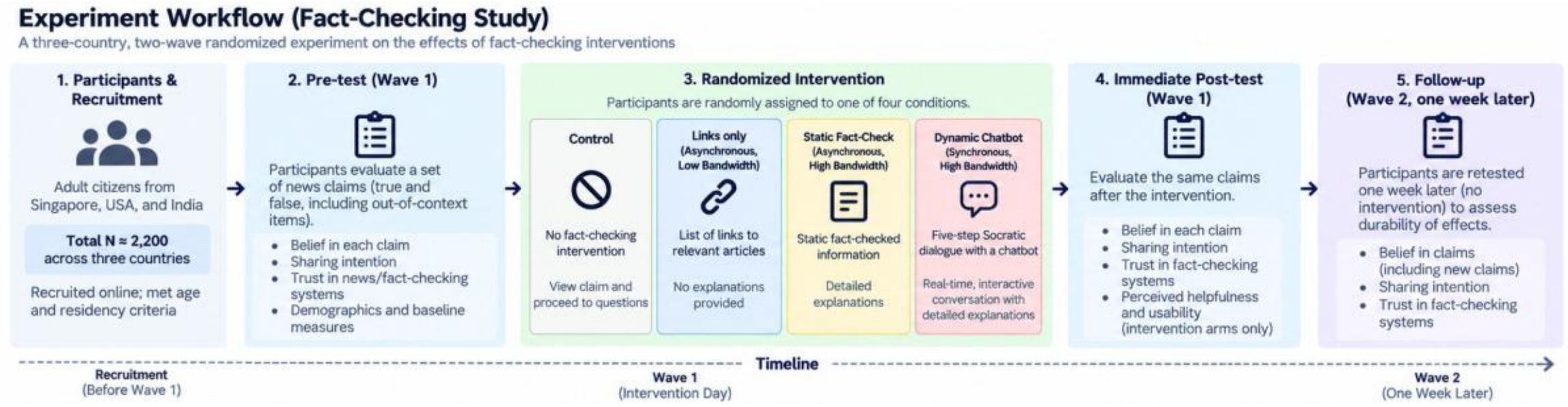

Fig. 2. Experiment workflow: recruitment, Wave 1 pre-test, randomization to one of four conditions, Wave 1 post-test, and the Wave 2 follow-up one week later.

Our study measures both constructs by manipulating *channel affordances* – properties of how a correction is delivered, rather than the specific reasoning it contains – along two dimensions established in the fact-checking and media-effects literature: synchronicity (whether the correction unfolds as a real-time interaction or a one-shot delivery) and bandwidth (how much explanatory richness it carries). For our study, we deployed *Sniffer*, a chatbot fact-checking system that we benchmarked against other models on the out-of-context misinformation task in prior published work. Sniffer is integrated as part of a five-step Socratic dialogue asks a participant questions and gives them room to engage with a claim's image, source, and evidence before the system delivers a verdict, varying outputs by the affordances available to the participant. Figure 2 summarizes the overall procedure.

We compared the chat-based factchecking treatment against a links-only correction (asynchronous, low-bandwidth), a static text explanation (asynchronous, high-bandwidth), and a no-intervention control, in a preregistered randomized experiment fielded across three countries (the United States, India, and Singapore). Chatbot sits at the high-synchronicity, high-bandwidth end of the same affordance scale Links and Static occupy at lower points. A two-wave setup lets us ask first, whether an intervention changes what people believe *right now* (object-level learning, testable at Wave 1), and further, whether it leaves anything behind once the chatbot itself is no longer present (metalearning, testable at Wave 2).

We pose the following research questions:

- **RQ1 (object-level learning):** Does a chatbot correction, varying in channel affordances relative to simpler correction formats, produce larger immediate gains in misinformation discernment and reduced sharing of false claims?
- **RQ2 (metalearning):** Does any condition's advantage persist to an unaided follow-up one week later – including on claims the chatbot never discussed – or does it decay once the system is removed?
- **RQ3 (affordance gradient):** If an advantage decays, does the amount of decay track the affordance level of the condition that produced it, or does it look like generic, condition-independent forgetting?

We report three findings, one per research question. First, at Wave 1, the Chatbot condition produced the largest immediate gains in discernment among the three interventions, and all three interventions (Links, Static, Chatbot) significantly reduced willingness to share false claims – clear object-level learning. Second, at Wave 2, none of these advantages persisted: two of three countries show a decline specifically in the Chatbot condition's relative standing, a pattern that survives checks for item-repetition and treatment-reinforcement confounds, and in India this decline appears on claims the chatbot never discussed at similar magnitude to claims it did, which is evidence against metalearning, and suggestive of a negative transfer effect. Third, within our own three conditions – a matched, single-exposure comparison that requires no external benchmark – the size of this decline tracks the affordance level of the condition: weakest for Links (asynchronous, low-bandwidth), strongest for Chatbot (synchronous, high-bandwidth), with Static in between. We interpret this through trust-calibration theory's account of engagement mechanisms (Section 2.2), and return to it directly in the Discussion.

## 2 Related Work

### 2.1 Out-of-Context Images: Why They Persuade and What Detects Them

Multimodal disinformation is rated more credible than the same claim in text alone, and although textual rebuttals reduce its credibility, they do not erase the advantage the image confers [15]. An authentic photograph, therefore, is an unusually effective vehicle for a false claim, because it builds legitimacy to the associated claim [25]. This is known as *the realism heuristic*: content that looks like a direct record of the world is credited with reliability [38]. Out-of-context (OOC) misinformation exploits this at minimum cost, as the falsehood lies entirely in how it is innocuously paired with a falsehood. Unsurprisingly, OOC misinformation is now the largest single category of visual misinformation that professional fact-checkers handle [17].

Detecting OOC misinformation is a different problem from text claim verification, which retrieves evidence and predicts whether it supports or refutes a claim [2, 36]: an OOC detector must reason jointly over the image and the caption. Early methods looked for inconsistencies within the image–text pair itself [4, 22]; later approaches added external evidence, retrieving related images and text from the web to test whether the claimed context is supported [1]. In this study, we integrate an OOC factchecking system, Sniffer [5], as part of a factchecking treatment. Sniffer first evaluates image-text consistency, and then retrieves contextual evidence before producing a judgment and an explanation (Section 3). However, like most factchecking systems evaluated on datasets, prior work has not established whether these OOC misinformation detectors are helpful to real humans encountering OOC misinformation, who receive a verdict and an explanation. Furthermore, no prior study of any OOC detector has examined what a user retains after it has explained a claim to them.

### 2.2 Channel Affordances of a Correction

Factchecking systems are increasingly available to help people verify the misinformation they encounter, and they differ in what they afford the user. The trust calibration framework [18] distinguishes *interpretability affordances* (e.g., source citations, uncertainty cues, which make independent evaluation *possible*) from *engagement mechanisms* (e.g., cooling-off periods, dialogue scaffolds, which make evaluation *happen* by building it into the interaction). The media-effects literature maps these onto two properties of a channel. Bandwidth – the capacity to carry multiple cues, elaboration, and personalization at once [14] – serves interpretability, and richer corrections are more effective ones [32]. Synchronicity – an exchange contingent on the user's own turns in real time – serves engagement: it raises perceived credibility through the interaction itself, whether or not the content improves [39], and allows the clarification and personalization a one-shot correction cannot [21, 27].

The framework also predicts that engagement bought this way is shallow: fast-acting engagement mechanisms are a thinner layer of defense than the slower-building AI literacy that produces metalearning [18]. The mechanism is social. People apply social heuristics – trust, reciprocity, assumed good faith – to systems that communicate conversationally, even when they know they are talking to a machine [24], and dual-process accounts of persuasion predict that conversational warmth, a confident tone, and apparent thoroughness are read as peripheral cues to reliability, displacing the systematic evaluation that discernment requires [30]. The user comes away trusting the verdict rather than having rehearsed the procedure that produced it. Two recent experiments corroborate this expectation. In a preregistered study in the United States and India (N=894), participants who received political information through a chatbot reported greater confidence and larger perceived knowledge gains than those who received the same content as a static list, yet scored *lower* on misinformation discernment [19]. Rani et al. [33] found that a month-long persuasive AI news-verification dialogue (N=67 completers) raised accuracy while it was available and, four weeks later without the system, left it below where it had started. Neither varied how the correction was delivered, and neither used a system validated on the detection task itself.

Chatbot-delivered fact-checks are increasingly how people encounter corrective information at all [12], and a chatbot is by default synchronous and high-bandwidth; but the same correction can be delivered with fewer affordances. Our conditions span that range: Control (no correction), Links (asynchronous, low-bandwidth), Static (asynchronous, high-bandwidth), and Chatbot (synchronous, high-bandwidth, with a Socratic turn-taking structure, built on Sniffer's detection architecture; Section 3).

### 2.3 Do Corrections Last, and Where?

Interventions against misinformation span a range from one-shot corrections, through inoculation, to education that aims to build a general skill, and what is known about whether each lasts is the baseline for our metalearning test. One-shot corrections can last longer than often assumed: fact-checks fielded simultaneously in Argentina, Nigeria, South Africa, and the United Kingdom improved belief accuracy, and the improvement was still detectable two weeks later [31]. Inoculation, whether as a game [8, 34] or as prebunking messages [35], builds resistance that persists for weeks after a single exposure, across cultures. Education in the fact-checker's own procedure – lateral reading, leaving a source to consult others rather than reading deeply within it [23, 41], taught as SIFT [11] – improves source evaluation. But the same literature documents decay at every level. Fact-check effects on COVID-19 misperceptions in the United States, Great Britain, and Canada were gone within weeks [10]; literacy gains erode without reinforcement [9, 20]; and an hour-long, in-person media-literacy training in India, among the most intensive interventions yet fielded, produced no measurable improvement in the ability to identify misinformation [7]. Brief accuracy nudges [28, 29] generalize across sixteen countries [6], but are designed to act in the moment, not to teach.

Two gaps in this evidence shape our design. First, almost all of it comes from single-country studies, yet country-level resilience to disinformation varies systematically with institutional trust, the fragmentation of the news environment, and the regulatory response [16], which are vastly different in the contexts we study: the United States, India, and Singapore. Second, every durable effect above comes from a static or educational format; whether a conversational, AI-mediated correction leaves anything behind after a comparable gap of days to weeks, with no repeated exposure, is untested – and Section 2.2 gives reason to expect that it does not. This paper addresses both gaps at once, testing an OOC-detection chatbot built the way one would actually be deployed against lower-affordance corrections in three countries, with a one-week unaided follow-up.

## 3 The Sniffer System

The Dynamic Chatbot condition is built around the detection architecture of Sniffer [5], a multimodal large language model developed specifically for *explainable out-of-context (OOC) misinformation detection*: cases where the image itself is authentic but has been paired with a caption that misrepresents what it depicts, which is also the specific kind of misinformation used in every stimulus item in this study. We chose Sniffer over a general-purpose fact-checking pipeline for two reasons. First, its detection task is a direct match to our materials: OOC misinformation is deliberately hard to catch because the visual content contains no manipulation for a detector (or a person) to find, and the deception depends entirely on cross-referencing the caption against external context. Second, Sniffer is independently validated at this task. Table 1 reproduces its published benchmark comparison on NewsCLIPpings [5], the largest OOC misinformation detection benchmark: Sniffer outperforms every baseline tested, including detectors trained from scratch and other pretrained multimodal pipelines, and its underlying InstructBLIP backbone alone – without Sniffer's task-specific tuning – classifies at only 47.4% accuracy, little better than chance, confirming that the improvement comes from Sniffer's OOC-specific design rather than from the base model's general capability. Beyond detection accuracy, Sniffer's explanations were validated with human evaluators: among OOC items users initially judged as real, 87% were correctly recategorized as fake after reading Sniffer's explanation, and for items already correctly judged as fake, the explanation increased users' stated confidence in that judgment for 42% of cases [5].

Table 1. Out-of-context misinformation detection accuracy (%) on NewsCLIPpings, reproduced from Anonymous [5]. All: overall accuracy. Fake: accuracy on out-of-context (fake) samples. Real: accuracy on not-out-of-context (real) samples.

| Method | All | Fake | Real |
|---|---|---|---|
| SAFE | 52.8 | 54.8 | 52.0 |
| EANN | 58.1 | 61.8 | 56.2 |
| VisualBERT | 58.6 | 38.9 | 78.4 |
| CLIP | 66.0 | 64.3 | 67.7 |
| DT-Transformer | 77.1 | 78.4 | 75.6 |
| CCN | 84.7 | 84.8 | 84.5 |
| Neu-Sym detector | 68.2 | – | – |
| **Sniffer** | **88.4** | **86.9** | **91.8** |

Sniffer detects OOC misinformation via two complementary checks – *internal checking* (does the image content itself contradict the caption?) and *external checking* (does retrieved outside context support or contradict the claim?) – and integrates both into a single explained judgment.

We adapted this two-part checking logic into an interactive, participant-facing dialogue rather than a one-shot backend classification. Sniffer is embedded as a web-based chatbot within the Qualtrics survey instrument (Figure 7). It presents a participant with one of six pieces of multimodal misinformation – an image or claim paired with a manipulated or out-of-context caption – and, in the Chatbot condition, opens with "Hello, I am a fact-checking assistant" and runs the participant through a scripted five-step Socratic dialogue that mirrors Sniffer's internal/external checking split: (1) checking text-image consistency, (2) checking text-evidence consistency (internal checking, steps 1–2), (3) verifying the claim's source, (4) evaluating the surrounding evidence (external checking, steps 3–4), and (5) summarizing the findings by integrating both (step 5). At each step the assistant asks the participant a question about the claim and incorporates their response before advancing, rather than issuing an immediate verdict; only after all five steps are complete does it state and justify a final assessment (true, false, or misleading). Two real example conversations from participant logs, spanning two of the three countries, are reported in Table 3 (Appendix).

For deployment at scale across three countries in a live, multi-turn conversational survey instrument, we implemented this checking logic as a scripted dialogue served by a general-purpose LLM (Meta's LLaMA-3.1-70B-Versatile via Groq's inference-only API) which offered an upgrade in terms of conversational ability. Groq's inference API does not retain or reuse submitted prompts for model training; all interaction logs, participant metadata, and experimental condition assignments are stored on university-owned AWS infrastructure under the study's IRB protocol, with no data shared with third-party platforms.

Chatbot-condition compliance, defined as at least one logged conversational exchange, was high and consistent across all three countries: 362/369 (98.1%) in the U.S., 377/384 (98.2%) in India, and 283/284 (99.6%) in Singapore. Across all three countries, chatbot-condition participants exchanged 12,912 messages across 3,823 logged conversations.

## 4 Method

### 4.1 Design and Participants

As seen in Figure 2, our preregistered, four-arm randomized experiment (Control, Links-only, Static explanation, Dynamic Chatbot) had a two-wave structure: an immediate post-treatment assessment (Wave 1) and an unaided follow-up assessment roughly one week later (Wave 2). Participants were recruited via Qualtrics panels in the United States, India, and Singapore. Fielding windows differed by country: the U.S. Wave 1 sample was collected October 16–30, 2024, in the weeks before the 2024 U.S. presidential election; India's Wave 1 sample was collected January 15–30, 2025; and Singapore's Wave 1 sample was collected January 7–February 20, 2025. Only the U.S. fielding window coincides with a comparably salient national election period; we treat cross-country comparisons as tests of the same intervention design under differing political-salience conditions, not as a controlled replication of identical contextual stakes, and return to this in Section 7.1.

Wave 1 randomized approximately one quarter of participants to each condition in every country (USA: 357 Control, 386 Links, 390 Static, 369 Chatbot; India: 397 Control, 390 Links, 389 Static, 384 Chatbot; Singapore: 304 Control, 311 Links, 321 Static, 284 Chatbot; full randomization table in the SI Appendix). Pooled across countries, the Wave 1 analysis sample used for the primary discernment and sharing outcomes totals approximately 2,200 participants (USA $n \approx 726$, India $n \approx 890$, Singapore $n \approx 610$, per-cell sample sizes vary slightly by outcome due to item non-response; exact per-comparison $n$, degrees of freedom, and full test statistics for every outcome in Figure 3 are reported in the SI Appendix). Wave 2 re-contacted a subset of the Wave 1 sample; linkage rates and Wave 2 sample sizes are reported in Section 5.2.

*Ethics, consent, and debriefing.* The study protocol was reviewed and approved by our institution's Institutional Review Board. Before Wave 1, participants read an information sheet describing the task (assessing the accuracy of a series of news headlines and images), its expected duration, the possibility of discomfort from controversial or misleading content, their right to withdraw at any time without giving a reason, and the reimbursement provided through the panel, and gave informed consent. No identifiable data were collected; interaction logs are stored as described in Section 3. Disclosure of the false items was staged so as not to contaminate the follow-up. In the three treatment conditions, every false item was accompanied at the point of correction by its original caption and a link to the source article, so participants learned during the study which images had been taken out of context; treatment participants were also told, before the post-treatment measures, that they had seen "many true and false news" items.

At the end of Wave 2, all participants, including those in Control, received a written debriefing that identified each false item, explained that the images were authentic but their captions had been altered for the study, and repeated the investigator's contact details. Debriefing was withheld until after Wave 2 so that the follow-up measured what participants retained rather than a second correction. Participants gave informed consent at the start of Wave 1, were compensated by the panel provider at its standard rate, and at the end of each wave were debriefed with the correct caption for every false item they had seen, including Control participants, who received no correction during the study. The design, primary outcomes, and analysis plan were preregistered on Open Science Foundation.[1]

Table 2 summarizes the Wave 1 sample by country. The three national samples differ substantially in age profile (USA skews oldest, with 35.6% aged 65+; India skews youngest, with 40.1% aged 25–34; Singapore is roughly even across the 25–64 range) and, for Singapore, in income level relative to its own country-specific bracket scale; gender is close to balanced in the USA and India, less so in Singapore (56.5% male). We verified that random assignment produced balanced groups: one-way ANOVAs (age, income, cognitive reflection) and chi-square tests (gender) comparing the four conditions within each country found no significant imbalance on any covariate in any country (all $p > .11$; full test statistics in the SI Appendix).

Table 2. Wave 1 sample demographics by country. Age and income are country-specific ordinal brackets (not on a common scale across countries; see SI Appendix for exact bracket definitions and the full age-bracket breakdown). Cognitive reflection is a 0–1 factor score.

| Country | $N$ | % Female | Mean age bracket | Mean income bracket | Cog. reflection |
|---|---|---|---|---|---|
| USA | 1502 | 49.7% | 4.46 / 6 | 3.17 / 6 | 0.14 |
| India | 1344 | 49.8% | 2.29 / 6 | 2.92 / 6 | 0.18 |
| Singapore | 1220 | 43.4% | 3.77 / 6 | 5.79 / 11 | 0.18 |

### 4.2 Materials and Conditions

All participants viewed six pieces of multimodal fake news (an image plus a false or misleading claim about it) drawn from a shared stimulus pool, verified against third-party fact-checks; at least one stimulus (concerning a Sentosa, Singapore oil-spill claim) was country-specific, and we treat the stimulus pool as matched-but-not-strictly-identical across countries rather than a single fixed instrument. Participants were randomly assigned to one of four conditions varying in synchronicity and explanatory depth:

- **Control**: no fact-checking; exposure to unrelated news content.
- **Links-only** (asynchronous, low-bandwidth): source links shown without further explanation.
- **Static explanation** (asynchronous, high-bandwidth): a written fact-check with supporting explanation.
- **Dynamic Chatbot** (synchronous, high-bandwidth): the scripted five-step Socratic dialogue described in Section 3.

### 4.3 Measures

We measure three outcomes at both Wave 1 (RQ1) and Wave 2 (RQ2, RQ3), plus a set of covariates used for balance checks and covariate-adjusted models.

[1](https://osf.io/v3uz4/overview?view_only=11136fbb433e4b99aa98fd4b85ce4f21)

**Misinformation discernment** (primary outcome) is the standard metric in the misinformation literature for separating genuine belief accuracy from a general tendency to rate everything as true or false [28]: because it uses both a true-claim and a false-claim rating rather than either alone, it cannot be inflated by a participant who simply rates every claim as accurate. We adapt it to our per-item, two-sub-rating design: it is scored per item on a 1–5 scale (higher = more accurate), using both sub-ratings collected for each stimulus so that higher always indicates greater accuracy regardless of the item's true veracity – for items labeled false, discernment is the average of (6 − truthful rating) and the fake rating; for items labeled true, it is the average of the truthful rating and (6 − fake rating). A participant's discernment score is the mean across the items they answered. This is our operationalization of both object-level learning (Wave 1, on items the chatbot discussed) and metalearning (Wave 2, on items it did not).

**Misinformation sharing** (primary outcome) matters because belief accuracy and willingness to spread a claim are theoretically and empirically separable [28]: an intervention could correct belief without changing sharing behavior, or vice versa, so both must be measured rather than assuming one implies the other. It is each participant's self-reported willingness to share the false-claim items specifically (1–5; lower is the hypothesized beneficial direction). We do not include true items in this measure, since willingness to share a claim one was never told is false is not a measure of misinformation sharing.

**Confidence** (secondary, exploratory outcome) is self-reported confidence in one's accuracy judgment, averaged across all items (true and false), with no directional hypothesis assumed. We include it because trust-calibration accounts of chatbot-mediated correction (Section 2) treat well-calibrated confidence as a goal in its own right [18]: a system that raises accuracy while leaving users overconfident, or appropriately less confident after encountering counter-evidence, is a different outcome than one that raises accuracy alone.

**Covariates**: cognitive reflection (a single-factor score derived from a battery of CRT-style items, identical across countries), age, gender, and household income (each derived from country-specific response categories). A political-ideology/party-identification measure is available for the U.S. and India but not Singapore, whose questionnaire instead includes a PAP-vs-WP affective feeling-thermometer item that is not a comparable ideology measure; we omit it from Singapore's covariate-adjusted models rather than substitute a non-equivalent proxy.

### 4.4 Analytic Strategy

For the primary Wave 1 outcomes (discernment, sharing) we report pairwise two-sample $t$-tests of each treatment against Control, pooled across countries and separately within each country, together with Cohen's $d$. As a robustness check, we additionally fit item-level models – crossed random-effects models (`discernment ~ treatment` + (1|participant) + (1|item)) and OLS with item fixed effects and participant-clustered standard errors – using every item observation rather than one averaged score per participant. Covariate-adjusted models add cognitive reflection, age, gender, income, and (where available) political ideology as controls.

For persistence, we estimate difference-in-differences models comparing each treatment's own Wave 1 → Wave 2 change in discernment against Control's own change, rather than simply re-testing whether a treatment effect remains individually significant at Wave 2 – a design that directly tests for *decay* rather than mere continued detectability. Because Wave 2 assigns participants to a second, independently randomized round of fact-checking support partway through the session, we restrict this comparison to items answered *before* that second randomization ("block 1"), so that any Wave 2 difference cannot be confounded by a fresh Wave 2 treatment. We report two versions of the Wave 1 baseline (the average across all Wave 1 items a participant answered, and their last false item specifically) and, as an independent third check, a "block 2" analysis using Wave 2 items answered *after* the second randomization, testing whether Wave 1's original treatment predicts these entirely later outcomes once Wave 2's own treatment is controlled for. We report achieved statistical power for each comparison post hoc, given the number of results in the persistence analysis that fall in a $p = .08$–$.24$ range where the distinction between a genuinely small effect and an underpowered test matters for interpretation.

## 5 Results

### 5.1 Immediate Effects (Wave 1)

We report pooled-across-countries estimates as primary throughout this section, each backed by an item-level OLS/mixed-effects model as a robustness check; per-country breakdowns are directionally consistent with the pooled estimates in every case (never reversed in sign) but are individually underpowered (25–29% achieved power; Section 5.2.2) and are reported in full in Appendix Tables 9–11 and Figure 8, not repeated here.

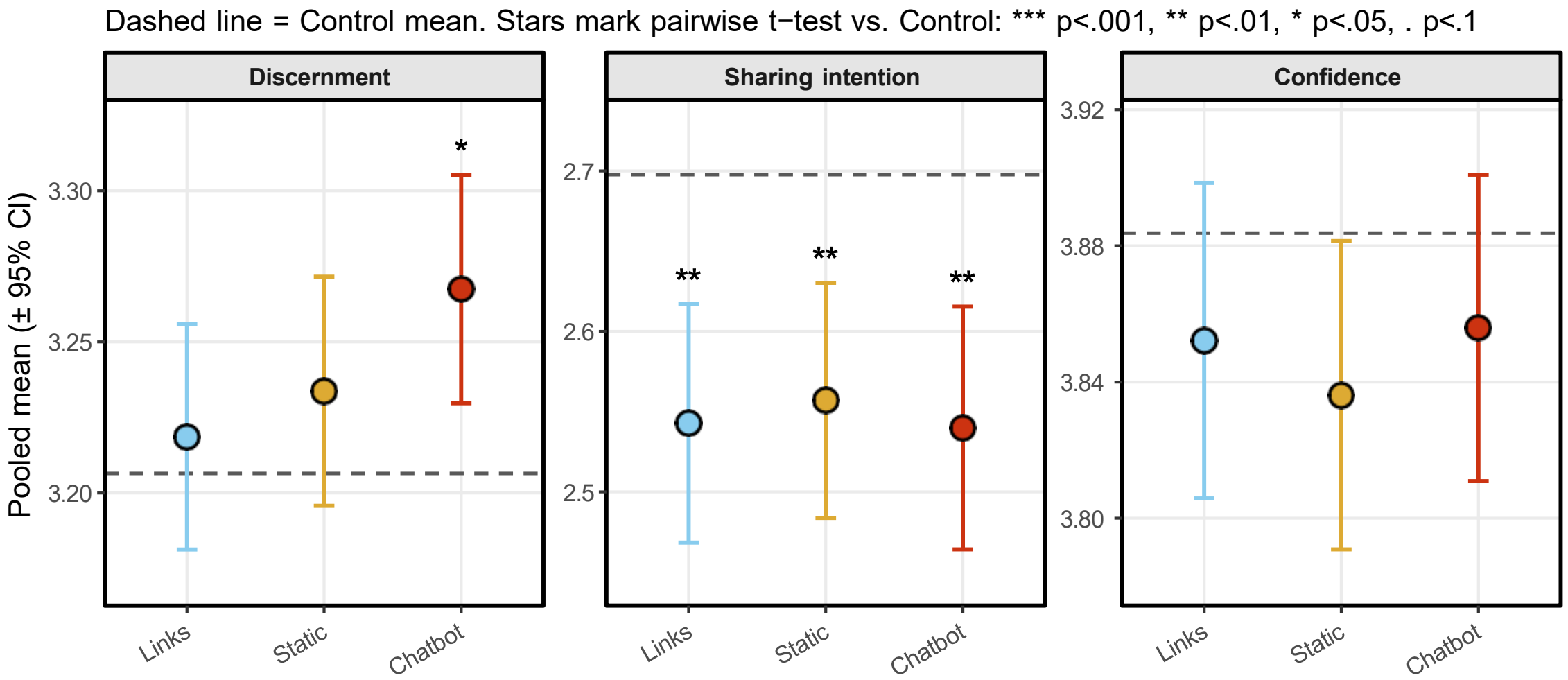

Fig. 3. Wave 1 outcomes by condition, pooled across countries, with 95% CIs. Discernment: higher is more accurate. Sharing intention: lower is less willing to share false claims. Confidence: no directional hypothesis. Per-country breakdown in Appendix Figure 8.

Pooled across all three countries, the Chatbot condition produced a significant improvement in discernment relative to Control ($t = 2.27$, $p = .023$, $d = 0.097$); Static ($d = 0.042$, $p = .312$) and Links ($d = 0.019$, $p = .650$) did not reach significance (Figure 3). An item-level crossed random-effects model and an OLS-with-item-fixed-effects/clustered-SE specification, using every item observation rather than one averaged score per participant, corroborate this: both estimate the pooled Chatbot effect at +0.073 ($p = .006$ and $p = .005$, respectively), with Static and Links again non-significant. Covariate-adjusted OLS models (cognitive reflection, age, gender, income, and, where available, political ideology; full coefficients in Appendix Table 13) show the same direction, with cognitive reflection and age emerging as strong, consistent predictors of discernment independent of treatment.

The sharing outcome is the more robust of the two primary results: pooled across countries, *all three* interventions significantly reduced willingness to share false claims (Links $d = -0.120$, $p = .005$; Static $d = -0.109$, $p = .010$; Chatbot $d = -0.122$, $p = .004$; Figure 3). Item-level models corroborate this with tighter estimates from using every item observation: all three treatments remain significant in both the mixed model and the OLS specification (Links $p$ = .008/.029, Static $p$ = .024/.059, Chatbot $p$ = .0017/.0022). Covariate-adjusted models show the same direction (full coefficients in Appendix Table 14), with cognitive reflection and age again strong, consistent predictors of lower sharing intention independent of treatment. Appendix Table 10 shows this pattern holding in India and Singapore for most or all conditions but not the USA, where sharing did not move significantly under any condition – notably the one country whose fielding window coincided with an active national election, a possible ceiling or context-salience effect we return to in Section 7.1.

Confidence, an exploratory secondary outcome, showed no significant pooled effect for any condition (Figure 3). One country-specific result is worth flagging even though it does not survive to the pooled estimate: India's Chatbot condition showed a significant confidence *decrease* relative to Control ($d = -0.140$, $p$ = .037; full per-country results in Appendix Table 11), replicated in an independent item-level model ($p$ = .041). We read this as a plausible calibration effect – chatbot users appropriately less certain after encountering counter-evidence – rather than a cause for concern, but it is a single significant cell among many null ones and should not be over-interpreted.

### 5.2 Fading at One-Week Follow-Up

Wave 1 and Wave 2 responses were linked at the individual level via each country's own recontact identifier (USA: 800/800 linked; India: 726/815, 89%) or, for Singapore, via a dedicated third-party crosswalk file (520 linked, deduplicated from raw exports that contained genuine duplicate rows), for a pooled linked sample of $N$ = 2,142. We estimate difference-in-differences (DiD) models: each treatment's own Wave 1 → Wave 2 change in discernment, relative to Control's own change, using only items measured before Wave 2's second, independently randomized round of fact-checking support.

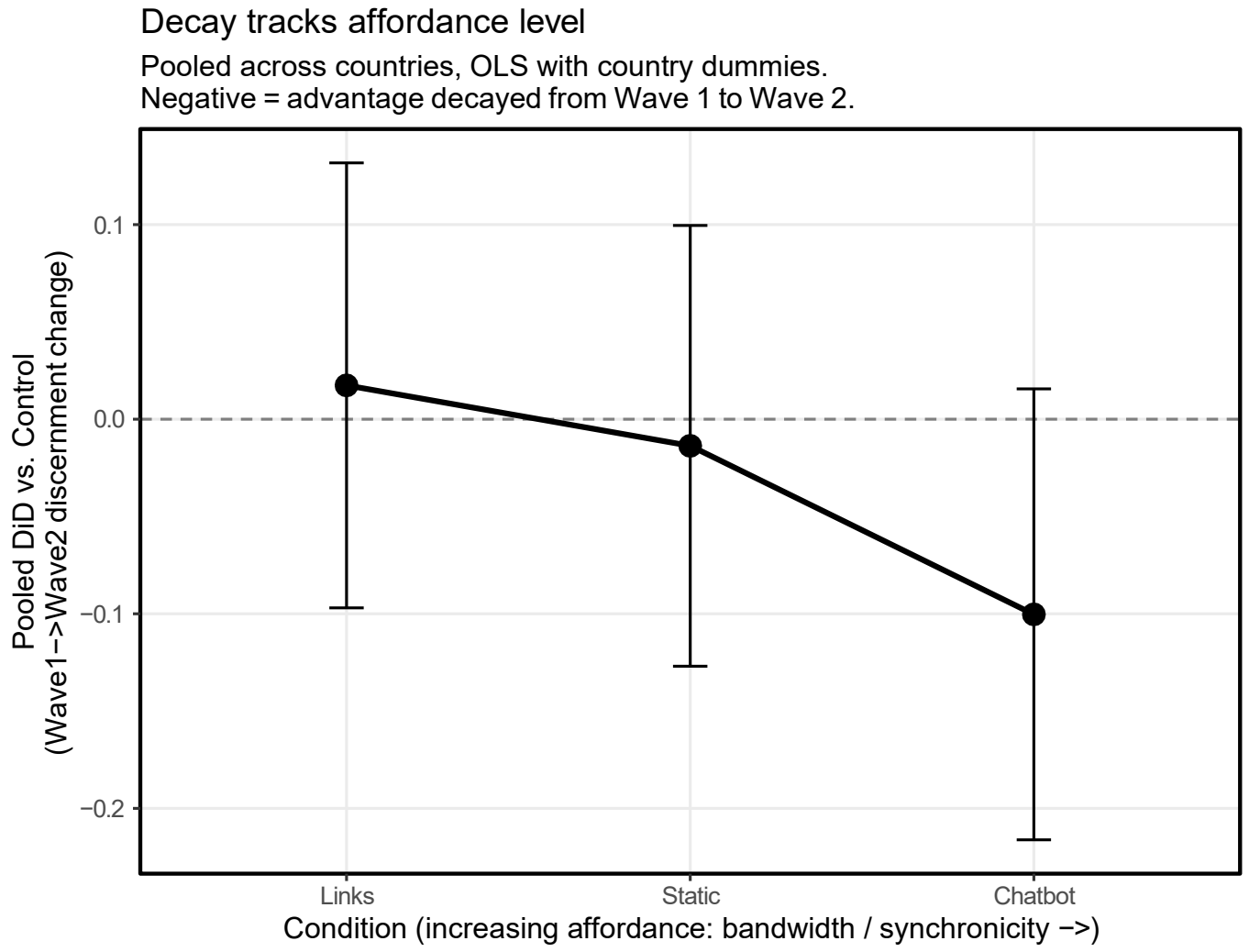

Fig. 4. Pooled Wave 1→Wave 2 difference-in-differences in discernment (OLS with country dummies; averaged-item baseline), plotted against affordance level. Negative values indicate the condition's relative advantage decayed.

No condition shows a significant *positive* pooled persistence effect. Links (+0.017, $p$ = .77) and Static (−0.014, $p$ = .81) are flat; Chatbot shows a decline that reaches the edge of conventional significance once pooled (−0.100, $p$ = .090; simple pooled $t$-test without the country covariate: $t = -1.95$, $p = .052$) – see Section 5.2.2 for the full model and the power case for reporting this pooled estimate as primary. This pattern – null for Links and Static, a marginal-to-significant decline concentrated in Chatbot – is consistent across every persistence test we ran (Figure 4). Two per-country results are individually well-powered and significant on their own: India's Chatbot condition under the last-item baseline ($d = -0.378$, $p < .001$, 92% power) and Singapore's Chatbot condition on repeated items specifically ($d = -0.448$, $p = .015$, 69% power; Section 5.2.1). USA shows no consistent treatment signal under any specification; Section 5.2.1 shows this is because USA's apparent persistence signal is fully explained by a generic item-repetition effect. Full per-country numbers for every baseline (averaged-item, last-item) and country are in Appendix Table 15.

A third, independent test using Wave 2 items answered *after* the second randomization, comprising new items relative to Wave 1's overlapping pool, asks whether Wave 1's original treatment predicts these later outcomes, controlling for Wave 2's own treatment. India's Chatbot condition again shows a marginal negative coefficient ($-0.116$, $p = .086$), consistent with the block-1 results above; Singapore shows a clean null on this test ($-0.033$, $p = .78$); USA shows a marginal *positive* coefficient for Chatbot ($+0.417$, $p = .064$), the opposite direction from India, on a test where – as with block 1 – USA's own repetition and reinforcement checks (below) find no consistent treatment signal at all.

5.2.1 *Ruling out repetition confounds.* Because Wave 2's block-1 items partially overlap Wave 1's stimulus pool, any persistence estimate could reflect a generic repetition or testing effect (having seen a claim before) rather than a treatment-specific effect. We tested this directly by splitting each country's linked sample into participants whose Wave 2 item repeated one of their own Wave 1 items versus participants seeing a genuinely novel item at Wave 2.

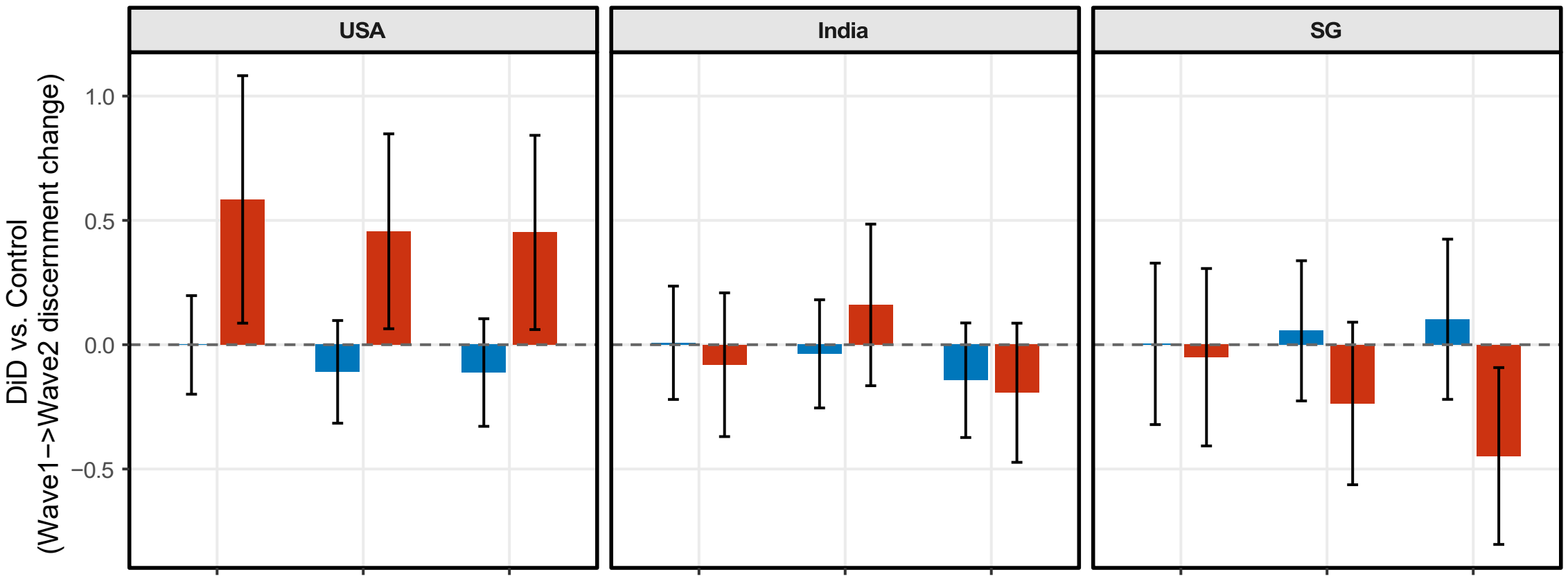

Fig. 5. Wave 1→Wave 2 discernment change (DiD vs. Control), split by whether the participant's Wave 2 item repeated one of their own Wave 1 items ("Repeat item") or was genuinely novel to them ("Novel item").

**USA's persistence signal is a repetition artifact.** As Figure 5 shows, all three conditions display an almost identical, statistically significant boost when the item repeats (Links $+0.584$, Static $+0.456$, Chatbot $+0.452$; all $p \approx .026$–$.027$) and sit at zero when it does not ($p$ = .30–.99). This is the signature of a generic testing effect that helps every condition equally, not a treatment effect, and explains why USA's per-country persistence estimates (Appendix Table 15) are noisy: repeat-item participants are a small minority diluted into a larger novel-item sample.

**India's decline is not a repetition artifact.** No condition in India shows anything resembling USA's uniform repeat-item boost; the Chatbot condition's negative direction appears at similar magnitude in *both* the repeat and novel subgroups (Figure 5), the opposite pattern from USA, where the effect was concentrated entirely in the repeat subgroup.

**Singapore's decline is repetition-*specific*, in the opposite direction from USA's.** Splitting Singapore's block-1 result (Figure 5) reveals a significant negative effect concentrated specifically in the Chatbot condition's repeat-item subgroup ($d$ = −0.448, $p$ = .015), with novel items trending slightly positive for every condition. This explains why Singapore's per-country estimate (Appendix Table 15) understates the effect: when a claim comes back around at Wave 2, Chatbot participants specifically perform *worse* than they did the first time, not merely no better.

We separately tested whether India's and Singapore's declines instead reflect a *treatment-repetition* effect – whether getting the same condition in both waves, versus a different one, matters – using block-2 data, where Wave 2's own (second) treatment assignment is observed. Neither country shows this: Singapore finds no significant difference between reinforced and switched participants for any condition ($p$ = .15–.71), and in India, Wave 2's second randomization assigns *every* participant to the Links condition, meaning no India participant who received Chatbot at Wave 1 ever receives it again at Wave 2 – so India's Chatbot decline cannot be a failure to reinforce the treatment, since the reinforced condition never occurs. Having ruled out both item-repetition and treatment-repetition as explanations, India's and Singapore's chatbot-specific declines remain the most credible signal in this analysis, arising through two distinct mechanisms (a general decline in India; a repetition-specific decline in Singapore) that converge on the same conclusion.

*5.2.2 Robustness check: pooling across countries.* Splitting the linked sample three ways by country leaves every per-country persistence test underpowered: 24–41% achieved power for the averaged-item baseline (Figure 6), the same pattern already seen at Wave 1, where the per-country discernment tests reach only 25–29% power against a pooled-significant effect. We pool India, Singapore, and USA's block-1 DiD data into a single OLS model, diff_discernment ∼ treatment + country, entering country as a set of dummy variables to absorb each country's own baseline shift in the discernment measure between block 1's Wave 1 and Wave 2 versions (different item sets, different practical response ranges) before estimating the treatment coefficients.

Note that an interaction model additionally allowing each treatment effect to vary by country does not fit significantly better ($F$ (6, 2130) = 0.55, $p$ = .77), which implies that the three per-country estimates are statistically consistent with one common effect split into three underpowered pieces, not with genuine cross-country heterogeneity that pooling would paper over.

Pooling roughly triples the effective sample size for the Chatbot comparison and raises achieved power from 24–41% to 49% (Figure 6), moving the estimate from deep in underpowered-null territory to the edge of conventional significance (−0.100, $p$ = .090; simple pooled $t$-test without the country covariate: $p$ = .052), while Links and Static remain null under the identical pooling (full model in Appendix Table 16). The same logic applies to the two individually well-powered per-country detections reported above: India's last-item result (92% power) and Singapore's repeat-item result (69% power) are real detections, not lucky noise, while the averaged-item persistence estimates in both countries (24–41% power) are directionally corroborating rather than independently confirmatory. Taken together – a pooled estimate at the edge of significance, two independently well-powered per-country detections in the same direction, and a null interaction test ruling out genuine cross-country heterogeneity – the achievable evidence, given the power this design has to offer, points toward a real, Chatbot-specific decline rather than noise.

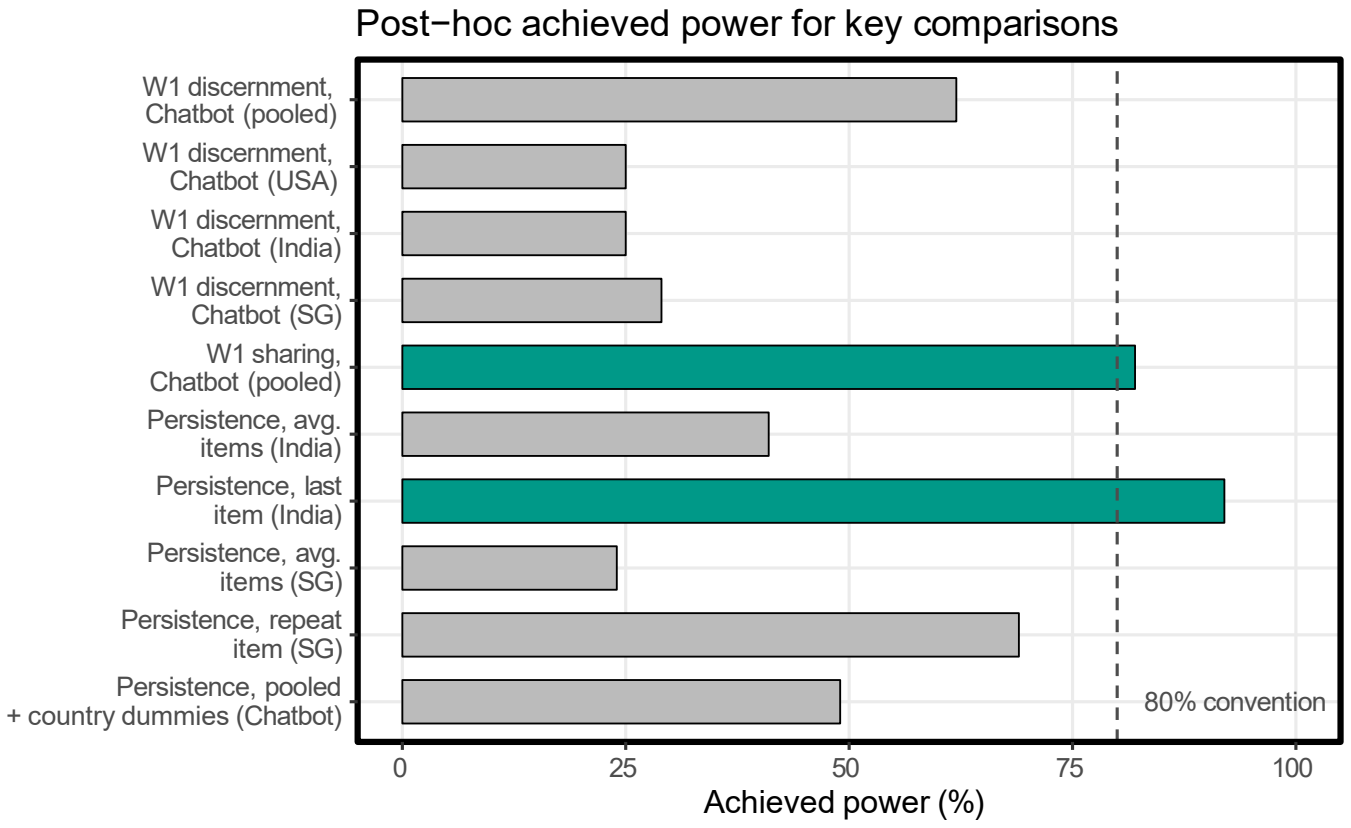

Fig. 6. Post-hoc achieved power for key comparisons, using observed effect sizes and sample sizes (unequal-$n$ two-sample tests). Bars reaching the 80% convention are highlighted.

## 6 Discussion

Our results suggest that chatbot-based fact-checking do not create enduring effects for misinformation discernment. What we find instead is that while the chatbot produced the largest immediate correction of any condition, a week later it demonstrated the largest relative decline. In India, the decline appears at similar magnitude on claims the chatbot never discussed, which resembles mild negative transfer onto novel claims. Below we read this pattern through the mechanism proposed in Section 2.2, and then return to the two gaps identified in Section 2.3. **Engagement without metalearning.** A synchronous, high-bandwidth dialogue supplies the social heuristics that dual-process models treat as peripheral routes to acceptance – responsiveness, a confident tone, apparent thoroughness [30]; accordingly, it is not surprising that participants extended social trust to Sniffer without paying attention to its procedure [24]. The Wave 2 decline, then demonstrates the learning that remains once the system is gone, when the participant has to deliver a verdict without a rehearsed procedure. Static delivers the same content without the interaction, and Links delivers only a pointer; each asks the participant to do more of the work, and each fades less. This result agrees with the trust-calibration prediction, that engagement mechanisms act fast but shallowly [18], now observed with the engagement level varied structurally rather than inferred from a single design. It also converges with observational evidence, that whether AI-mediated interaction builds capability depends on how the user engages with the assistant's teaching, not on whether teaching is offered [37]. Sniffer's dialogue offered the identical scripted procedure to every participant; however, that uniformity was not enough.

Two findings complicate a purely pessimistic reading: first, the reduction in sharing-intention reduction shows that a single chatbot correction can shift a stated behavioral intention, even where it leaves no metalearning. Second the decrease in confidence in India's chatbot-condition at Wave 1 is the opposite of the confidence-without-competence pattern that a conversational format produced in political information seeking [19]: here, the dialogue confronted participants with counter-evidence rather than summarizing content for them, and they became appropriately less certain. Whether a chatbot inflates or calibrates confidence appears to depend on the goals and not the format of the conversation that participants have.

The interaction logs offer a small, unsolicited window onto the same mechanism. The chatbot's final turn summarized its findings and asked whether the participant wished to continue or close; 76 of 6,280 logged conversations contain a substantive reply, and most of these are verdict acceptances rather than reasoning: *"I will accept this as fact"* (man, 55–64, USA), *"i agree with the conclusion"* (woman, 55–64, USA), *"It is misleading now that I know. Close the conversation"* (woman, 25–34, USA). One India participant (man, 45–54) acknowledged the correction, yet dismissed it – *"the image can be different but the cause of poor indian workers are still the same"* – which illustrates how the object-level correction of the image can leave the underlying belief untouched. Others bypassed the procedure in the opposite direction, from prior belief: *"The image doesn't matter I already knew this was false before I saw the image"* (man, 65+, USA). A few reported the effort the design intended (*"I guess I had to think a little . . . I guess it's good to be slightly stimulated"* ; man, 55–64, USA) or asked how to keep using the system (*"is there your app available?"* ; man, 25–34, India). These are illustrations, as only fewer than two percent of conversations volunteered feedback. However, they offer useful insights into how the verdict was received, and what participants paid attention to.

**How interventions could endure.** In prior work, static fact-checks have produced belief accuracy gains still detectable two weeks later in four countries [31]; other fact-check effects have been gone within weeks [10]. Our Static and Links conditions fall in between – their Wave 1 advantages did not persist, but neither did they reverse. The Chatbot condition did reverse, relative to Control, and did so in two of three countries on a pattern that survives item-repetition and treatment-repetition checks. This replicates the direction of Rani et al. [33]' month-long finding with a one-week gap, a single exposure, a structurally different design, and a system validated on the detection task [5], which makes it unlikely that the earlier result was an artefact of persuasion tactics or of a weak underlying fact-checker. The India result is worth setting beside prior work that an hour-long, in-person media-literacy training in India produced no measurable improvement [7]: in a context where even intensive education did not transfer, a single conversation transferred negatively.

**Cross-country similarities and differences.** Our study contexts differ in institutional trust, news-environment fragmentation, and regulatory response [16]. The immediate, object-level findings of our work were context-sensitive: sharing intentions moved in India and Singapore but not in the United States, where fielding coincided with a presidential election, and confidence fell only in India. However, the fading did not vary. The findings suggest that the shallowness of an engagement mechanism is a property of the mechanism rather than of the context it is deployed in. However, as the per-country persistence tests are underpowered, they offer fertile grounds for future work.

### 6.1 Design Implications

A single chatbot correction, including one built around Socratic questioning, is well-suited to correcting belief about the claim in front of the user and to reducing the intention to share it. However, it should not be assumed to also produce metalearning that outlasts the conversation. Three implications follow. First, in trust-calibration terms, a system that wants transfer should invest in interpretability affordances – the source, the original caption, the reverse-image-search result, exposed so that the user can perform the check – rather than in engagement mechanisms that perform it for them [18]. Second, cognitive apprenticeship's sequence of modeling, scaffolding, and fading [13] suggests a concrete design target: a dialogue that models the procedure once, then progressively withholds steps and asks the user to supply them, across several claims spaced over time, rather than delivering the full scaffold every time. Third, evaluation should change. Sniffer's benchmark accuracy and its human-rated explanation quality [5] were both excellent, and neither predicted what participants retained. Therefore, an unaided follow-up on novel claims should be a standard test for any fact-checking chatbot that claims to help users, not only to help verdicts.

## 7 Conclusion

A chatbot built on a validated out-of-context detector, deployed to roughly 2,200 people in the United States, India, and Singapore, produced the largest immediate discernment gain of any condition and a significant reduction in stated sharing intentions – clear object-level learning – and, one week later, the largest relative decline, ordered by affordance level and, in India, extending to claims it never discussed. Chatbot engagement did not beget metalearning.

The study closes two gaps. Conversational, AI-mediated correction had not been tested for persistence; it now sits at the ephemeral end of the correction literature, with a fade steeper than that of static corrections of the same content. Second, prior literature had been single-country; across three media environments that differ in trust, fragmentation, and regulation, the immediate effects varied while the fade did not. The practical conclusion is that a single, richly interactive correction and metalearning are different design targets reached by different means: the first by engagement, the second by interpretability affordances, spaced practice, and a scaffold that is withdrawn. Fact-checking systems that claim the second should be built for it, and evaluated for it, with appropriately calibrated designs.

### 7.1 Limitations

Several limitations qualify these findings. First, fielding windows differed substantially across countries – only the U.S. sample was collected during a comparably salient national election period – so cross-country comparisons should be read as tests of the same intervention design under differing contextual salience, not as a controlled replication of identical stakes; this may plausibly explain why the U.S. is the one country where the sharing outcome did not move. Second, all outcomes are self-reported; we do not have a behavioral (e.g., actual sharing) measure of misinformation spread. Finally, the deployed dialogue was scripted and served through a third-party inference API, and a small number of participants remarked on the cost of that in the logs – on latency (*"for an award winning chatbot why are replies taking 2 mints long?"* ; Singapore) and on rigidity (*"this Chatbot doesn't understand simple English replies, just keeps saying the same thing"* ; man, 55–64, USA), which we had implemented through rules in the context window, so that conversations stay on track for factchecking. However, a slower or more repetitive interaction plausibly dampens engagement. Therefore, the Chatbot condition's Wave 1 advantage may understate what a more fluent system would produce, though there is no reason to expect it to change what persists.

## 8 LLM Usage Disclosure

We used LLMs for minor writing assistance, including grammar correction and language polishing. The core research ideas, methodology, experimental design, implementation, analysis, and conclusions were developed and carried out by the authors.

## References

[1] Sahar Abdelnabi, Rakibul Hasan, and Mario Fritz. 2022. Open-Domain, Content-Based, Multi-Modal Fact-Checking of Out-of-Context Images via Online Resources. In *Proceedings of the IEEE/CVF Conference on Computer Vision and Pattern Recognition (CVPR)*. 14940–14949.

[2] Rami Aly, Zhijiang Guo, Michael Schlichtkrull, James Thorne, Andreas Vlachos, Christos Christodoulopoulos, Oana Cocarascu, and Arpit Mittal. 2021. Feverous: Fact extraction and verification over unstructured and structured information. *arXiv preprint arXiv:2106.05707* (2021).

[3] Zeenab Aneez, Taberez Ahmed Neyazi, Antonis Kalogeropoulos, and Rasmus Kleis Nielsen. 2025. *India Digital News Report*. Technical Report. University of Oxford. https://reutersinstitute.politics.ox.ac.uk/our-research/india-digital-news-report.

[4] Shivangi Aneja, Chris Bregler, and Matthias Nießner. 2023. COSMOS: Catching Out-of-Context Misinformation with Self-Supervised Learning. In *Proceedings of the AAAI Conference on Artificial Intelligence*.

[5] Anonymous. 2024. Anonymous for peer review.

[6] Antonio A. Arechar, Jennifer Allen, Adam J. Berinsky, Rocky Cole, Ziv Epstein, Kiran Garimella, Andrew Gully, Jackson G. Lu, Robert M. Ross, Michael N. Stagnaro, Yunhao Zhang, Gordon Pennycook, and David G. Rand. 2023. Understanding and Combatting Misinformation Across 16 Countries on Six Continents. *Nature Human Behaviour* 7, 9 (2023), 1502–1513. doi:10.1038/s41562-023-01641-6

[7] Sumitra Badrinathan. 2021. Educative Interventions to Combat Misinformation: Evidence from a Field Experiment in India. *American Political Science Review* 115, 4 (2021), 1325–1341. doi:10.1017/S0003055421000459

[8] Melisa Basol, Jon Roozenbeek, and Sander van der Linden. 2020. Good News about Bad News: Gamified Inoculation Boosts Confidence and Cognitive Immunity Against Fake News. *Journal of Cognition* 3, 1 (2020), 2. doi:10.5334/joc.91

[9] Joel Breakstone, Mark Smith, Nadav Ziv, and Sam Wineburg. 2022. Civic preparation for the digital age: How college students evaluate online sources about social and political issues. *The Journal of Higher Education* 93, 7 (2022), 963–988.

[10] John M. Carey, Andrew M. Guess, Peter J. Loewen, Eric Merkley, Brendan Nyhan, Joseph B. Phillips, and Jason Reifler. 2022. The Ephemeral Effects of Fact-Checks on COVID-19 Misperceptions in the United States, Great Britain and Canada. *Nature Human Behaviour* 6, 2 (2022), 236–243. doi:10.1038/s41562-021-01278-3

[11] Mike Caulfield. 2019. SIFT (The Four Moves). Hapgood. https://hapgood.us/2019/06/19/sift-the-four-moves/, Accessed: 2026.

[12] Michael Chan. 2026. "Is This Fake News?" Examining the Antecedents of Generative AI Chatbot Use for Fact-Checking of News Online. *Journal of Broadcasting & Electronic Media* (2026), 1–18.

[13] Allan Collins. 1989. Cognitive apprenticeship: Teaching the crafts of reading, writing, and mathematics. *Knowing, learning, and instruction: Essays in honor of...* (1989).

[14] Richard L. Daft and Robert H. Lengel. 1986. Organizational Information Requirements, Media Richness and Structural Design. *Management Science* 32, 5 (1986), 554–571. doi:10.1287/mnsc.32.5.554

[15] Michael Hameleers, Thomas E. Powell, Toni G. L. A. Van Der Meer, and Lieke Bos. 2020. A Picture Paints a Thousand Lies? The Effects and Mechanisms of Multimodal Disinformation and Rebuttals Disseminated via Social Media. *Political Communication* 37, 2 (2020), 281–301. doi:10.1080/10584609.2019.1674979

[16] Edda Humprecht, Frank Esser, and Peter Van Aelst. 2020. Resilience to Online Disinformation: A Framework for Cross-National Comparative Research. *The International Journal of Press/Politics* 25, 3 (2020), 493–516. doi:10.1177/1940161219900126

[17] ImageWhisperer. 2026. Yearly Fact Check Intelligence Report. https://imagewhisperer.org/yearly-report.

[18] Kokil Jaidka and Mengxuan Cai. 2026. Why We Believe Chatbots: Trust Calibration as a Design Problem. *Frontiers in Psychology* 17 (2026), 1935527.

[19] Kokil Jaidka and Shaz Furniturewala. 2025. Chatbot-guided Search delivers Low-Relevance News and can exacerbate Gender Gaps in Political Knowledge. Preprint. doi:10.21203/rs.3.rs-7106857/v1

[20] Joseph Kahne and Benjamin Bowyer. 2017. Educating for democracy in a partisan age: Confronting the challenges of motivated reasoning and misinformation. *American educational research journal* 54, 1 (2017), 3–34.

[21] Makoto Kato, Ryen W. White, Jaime Teevan, and Susan Dumais. 2013. Clarifications and Question Specificity in Synchronous Social Q&A. In *Proceedings of the SIGCHI Conference on Human Factors in Computing Systems*. ACM.

[22] Grace Luo, Trevor Darrell, and Anna Rohrbach. 2021. NewsCLIPpings: Automatic Generation of Out-of-Context Multimodal Media. In *Proceedings of the 2021 Conference on Empirical Methods in Natural Language Processing (EMNLP)*. Association for Computational Linguistics, 6801–6817. doi:10.18653/v1/2021.emnlp-main.545

[23] Sarah McGrew. 2024. Teaching Lateral Reading: Interventions to Help People Read Like Fact Checkers. *Current Opinion in Psychology* 55 (2024), 101737. doi:10.1016/j.copsyc.2023.101737

[24] Clifford Nass and Youngme Moon. 2000. Machines and Mindlessness: Social Responses to Computers. *Journal of Social Issues* 56, 1 (2000), 81–103. doi:10.1111/0022-4537.00153

[25] Eryn J. Newman, Maryanne Garry, Daniel M. Bernstein, Justin Kantner, and D. Stephen Lindsay. 2012. Nonprobative Photographs (or Words) Inflate Truthiness. *Psychonomic Bulletin & Review* 19, 5 (2012), 969–974. doi:10.3758/s13423-012-0292-0

[26] Nic Newman, Amy Ross Arguedas, Craig T. Robertson, Rasmus Kleis Nielsen, and Richard Fletcher. 2025. *Reuters Institute Digital News Report 2025*. Technical Report. Reuters Institute for the Study of Journalism, University of Oxford. https://reutersinstitute.politics.ox.ac.uk/digital-news-report/2025.

[27] Katy E. Pearce and Pranav Malhotra. 2022. Inaccuracies and Izzat: Channel Affordances for the Consideration of Face in Misinformation Correction. *Journal of Computer-Mediated Communication* 27, 2 (2022), zmac004. doi:10.1093/jcmc/zmac004

[28] Gordon Pennycook, Ziv Epstein, Mohsen Mosleh, Antonio A Arechar, Dean Eckles, and David G Rand. 2021. Shifting attention to accuracy can reduce misinformation online. *Nature* 592, 7855 (2021), 590–595.

[29] Gordon Pennycook, Jonathon McPhetres, Yunhao Zhang, Jackson G Lu, and David G Rand. 2020. Fighting COVID-19 misinformation on social media: Experimental evidence for a scalable accuracy-nudge intervention. *Psychological science* 31, 7 (2020), 770–780.

[30] Richard E. Petty and John T. Cacioppo. 1986. The Elaboration Likelihood Model of Persuasion. In *Advances in Experimental Social Psychology*. Vol. 19. Academic Press, 123–205. doi:10.1016/S0065-2601(08)60214-2

[31] Ethan Porter and Thomas J. Wood. 2021. The Global Effectiveness of Fact-Checking: Evidence from Simultaneous Experiments in Argentina, Nigeria, South Africa, and the United Kingdom. *Proceedings of the National Academy of Sciences* 118, 37 (2021), e2104235118. doi:10.1073/pnas.2104235118

[32] Toby Prike and Ullrich K. H. Ecker. 2023. Effective Correction of Misinformation. *Current Opinion in Psychology* 54 (2023), 101712. doi:10.1016/j.copsyc.2023.101712

[33] Anku Rani, Valdemar Danry, Paul Pu Liang, Andrew Lippman, and Pattie Maes. 2026. Dialogues with AI Reduce Beliefs in Misinformation but Build No Lasting Discernment Skills. In *Proceedings of the 2026 CHI Conference on Human Factors in Computing Systems*. 1–26.

[34] Jon Roozenbeek and Sander van der Linden. 2019. Fake News Game Confers Psychological Resistance Against Online Misinformation. *Palgrave Communications* 5, 1 (2019), 65. doi:10.1057/s41599-019-0279-9

[35] Jon Roozenbeek, Sander Van Der Linden, and Thomas Nygren. 2020. Prebunking interventions based on "inoculation" theory can reduce susceptibility to misinformation across cultures. *Harvard Kennedy School Misinformation Review* 1, 2 (2020).

[36] Michael Schlichtkrull, Zhijiang Guo, and Andreas Vlachos. 2023. Averitec: A dataset for real-world claim verification with evidence from the web. *Advances in Neural Information Processing Systems* 36 (2023), 65128–65167.

[37] Yijia Shao, Dora Zhao, Vishakh Padmakumar, Jennifer Wang, and Diyi Yang. 2026. Human–AI Collaboration at Scale: Task Criticality, Agency, and Friction Across 250,000 Conversations. https://www.alphaxiv.org/abs/2608.human-ai-collaboration-at-scale.

[38] S. Shyam Sundar. 2008. The MAIN Model: A Heuristic Approach to Understanding Technology Effects on Credibility. In *Digital Media, Youth, and Credibility*, Miriam J. Metzger and Andrew J. Flanagin (Eds.). The MIT Press, Cambridge, MA, 73–100. doi:10.1162/dmal.9780262562324.073

[39] S. Shyam Sundar, Haiyan Jia, T. Franklin Waddell, and Yan Huang. 2015. Toward a Theory of Interactive Media Effects (TIME): Four Models for Explaining How Interface Features Affect User Psychology. In *The Handbook of the Psychology of Communication Technology*, S. Shyam Sundar (Ed.). Wiley Blackwell, 47–86. doi:10.1002/9781118426456.ch3

[40] We Are Social and Meltwater. 2024. Digital 2024: Global Overview Report. https://datareportal.com/reports/digital-2024-global-overview-report.

[41] Sam Wineburg and Sarah McGrew. 2019. Lateral reading and the nature of expertise: Reading less and learning more when evaluating digital information. *Teachers College Record* 121, 11 (2019), 1–40.

## A System Architecture

See Figure 7.

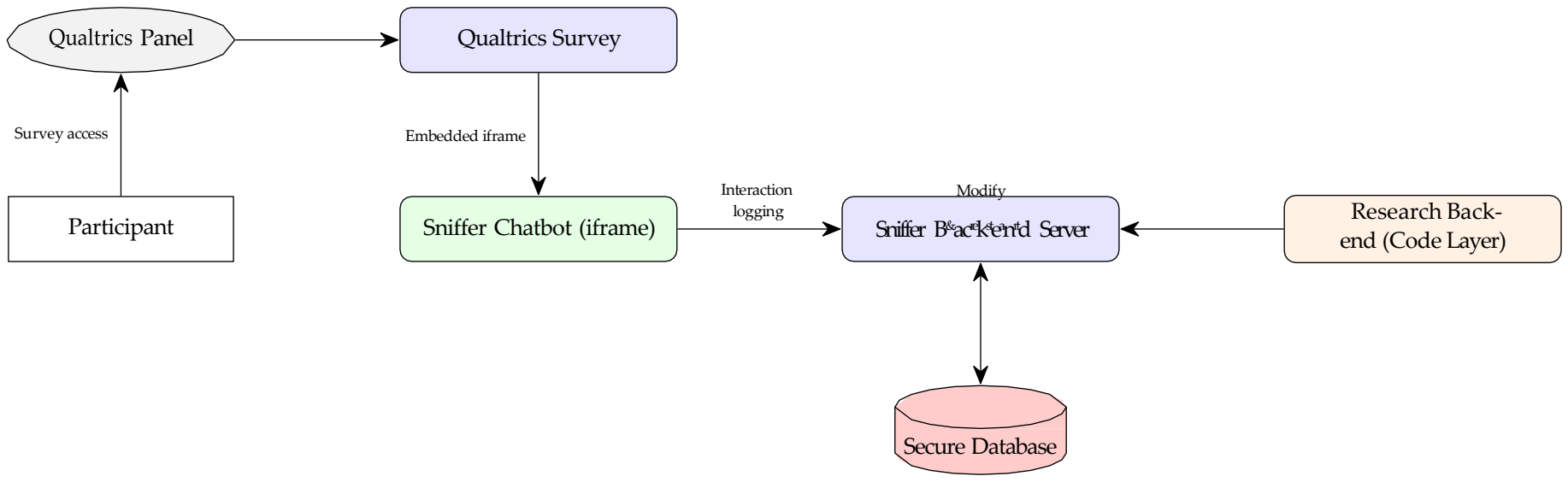

Fig. 7. System architecture of the Sniffer chatbot platform, embedded within Qualtrics for use in a controlled field experiment. Participant interactions are logged by the backend and stored on institution-owned infrastructure.

## B Example Conversations

See Table 3.

## C Outcome Scoring

### C.1 Misinformation discernment

Each of the six stimulus items is labeled, by design, as depicting either a false or a true claim. For every item a participant answered, we collected two sub-ratings on a 1–5 scale: how *truthful* they judged the claim to be, and how *fake* they judged it to be. Because these two ratings are not automatically aligned (e.g., a participant could rate a false claim as both moderately truthful and moderately fake), we compute a single discernment score per item that is symmetric and always oriented so that higher values mean greater accuracy, regardless of the item's true label:

$$\text{discernment (false-labeled item)} = \tfrac{1}{2}\left[(6 - \text{truthful rating}) + \text{fake rating}\right]$$

$$\text{discernment (true-labeled item)} = \tfrac{1}{2}\left[\text{truthful rating} + (6 - \text{fake rating})\right]$$

A participant's overall discernment score is the mean of this quantity across every item they answered (India's Qualtrics flow shows each participant only 2–3 of the six items via a block-randomizer; USA and Singapore show all six). Item-level, rather than only person-level, scores are retained so that item and participant random/fixed effects can be estimated separately (Table 12, Section F).

### C.2 Misinformation sharing

Sharing intention is the self-reported *-share* rating (1–5, higher = more willing to share) collected for false-labeled items only. We exclude true-labeled items from this measure because willingness to share a claim a participant was never told was false is not a measure of willingness to spread misinformation.

### C.3 Confidence

Confidence is the self-reported *-confident* rating (1–5), averaged across all items a participant answered, true and false alike, with no directional hypothesis assumed.

### C.4 Covariates

Cognitive reflection is derived identically across all three countries from a battery of Cognitive Reflection Test-style items. Age, gender, and income are drawn from each country's own demographic questions, using each country's own response-category structure (e.g., Singapore's income brackets are in SGD and more finely graded than the USA's USD brackets); these are therefore harmonized in *direction* and *scale position* across countries but not in the underlying currency or category boundaries. Political ideology/party-identification is available for the USA and India, using each country's own (non-equivalent) scale; Singapore's questionnaire does not ask a left-right or party self-placement question, only a PAP-vs-WP feeling-thermometer item, which we treat as measuring affective polarization rather than ideology and therefore exclude from Singapore's covariate models rather than substitute as a proxy.

## D Participants, Randomization, and Materials

### D.1 Sample sizes by country and condition (Wave 1)

USA's counts were tallied directly from the raw Wave 1 export's treatment-assignment column (the same method used for India and Singapore above); the Chatbot-condition count (369) matches the number used in the compliance calculation (Section E) and is independently verified against the interaction-log crosswalk. All four USA counts also match the number of USA participants who went on to answer at least one scored discernment item, i.e. no USA participant was randomized but never responded to any item.

### D.2 Fielding windows

Only the USA window overlaps a comparably salient national election period (the 2024 U.S. presidential election). We do not treat the three countries' fielding windows as contextually equivalent; see the main text's Limitations section.

### D.3 Sample demographics and covariate balance

Table 7 reports the full covariate-balance check summarized in the main text's Method section: one-way ANOVAs (age, income, cognitive reflection) and chi-square tests (gender) comparing the four Wave 1 conditions within each country. No covariate shows a significant imbalance in any country.

### D.4 Stimulus materials

All participants were shown six pieces of multimodal misinformation (an image or claim paired with a manipulated or out-of-context caption), verified against third-party fact-checks. At least one stimulus item used in the Singapore sample (a claim about an oil spill near Sentosa) was country-specific rather than shared across all three countries; we were not able to fully audit, from the materials available for this SI, which of the remaining five items are strictly identical across all three countries versus country-adapted variants of the same underlying claim. We report this as an open item rather than assert full cross-country stimulus equivalence.

## E Chatbot Compliance

Compliance is defined as at least one logged conversational exchange between a Chatbot-condition participant and the Sniffer system, matched via each participant's survey identifier to the interaction-log database.

An earlier pull of the interaction logs, frozen on 2025-02-02, understated Singapore's compliance (52.1%) because Singapore's Wave 1 fielding continued through February 20, 2025; participants whose sessions fell after the export cutoff had time-on-page durations on the fact-check step statistically indistinguishable from participants whose sessions were captured (medians of 56 vs. 54 seconds), consistent with real, uncaptured engagement rather than genuine non-compliance. A refreshed export (through June 13, 2025) resolved this, and Table 8 reports the corrected figures. Across all three countries, chatbot-condition participants exchanged 12,912 messages across 3,823 logged conversations.

## F Full Regression Tables

The main text presents Wave 1 and persistence results as figures (Sections 4.1 and 5.2 there); this section gives the full numeric tables behind each figure, for readers who want exact test statistics rather than a visual summary.

### F.1 Wave 1 primary outcomes: pairwise $t$-tests

Behind Figure 1 of the main text (Wave 1 outcomes, pooled). The main text reports pooled estimates only; this section gives the full per-country breakdown, plotted in Figure 8 and tabulated below. Per-country estimates are directionally consistent with the pooled estimates throughout – never reversed in sign – but individually underpowered (25–29% achieved power for discernment; Section 5.2.2 of the main text), which is why the main text does not treat non-significant per-country cells as evidence against the pooled finding.

### F.2 Discernment: item-level secondary models

### F.3 Discernment: covariate-adjusted models

### F.4 Sharing: covariate-adjusted models

### F.5 Persistence: difference-in-differences

Behind the pooled-estimate figure in the main text's Results (Section 5.2); this section gives the full per-country breakdown and the pooled model behind it.

Fig. 8. Wave 1 outcomes by condition and country (pooled and per-country), with 95% CIs. Discernment: higher is more accurate. Sharing intention: lower is less willing to share false claims. Confidence: no directional hypothesis.

### F.6 Pooled persistence model

Behind the "Robustness check: pooling across countries" subsection in the main text (Section 5.2.2). OLS, block-1 discernment difference score (Wave 2 − Wave 1) regressed on treatment (reference: Control) with country entered as a covariate, pooling India, Singapore, and USA's linked participants.

$N$ = 2142. A treatment × country interaction model does not fit significantly better ($F$ (6, 2130) = 0.55, $p$ = .77), i.e. there is no evidence the treatment effects differ by country – the statistical basis for reporting the pooled estimate as a single summary rather than as three independent per-country tests. The two country coefficients are large and highly significant because they capture each country's own baseline shift in the discernment measure between block-1's Wave 1 and Wave 2 versions (e.g. different item sets, different response scales in practice), not a treatment effect; they are included purely to absorb that baseline variation before estimating the treatment coefficients above.

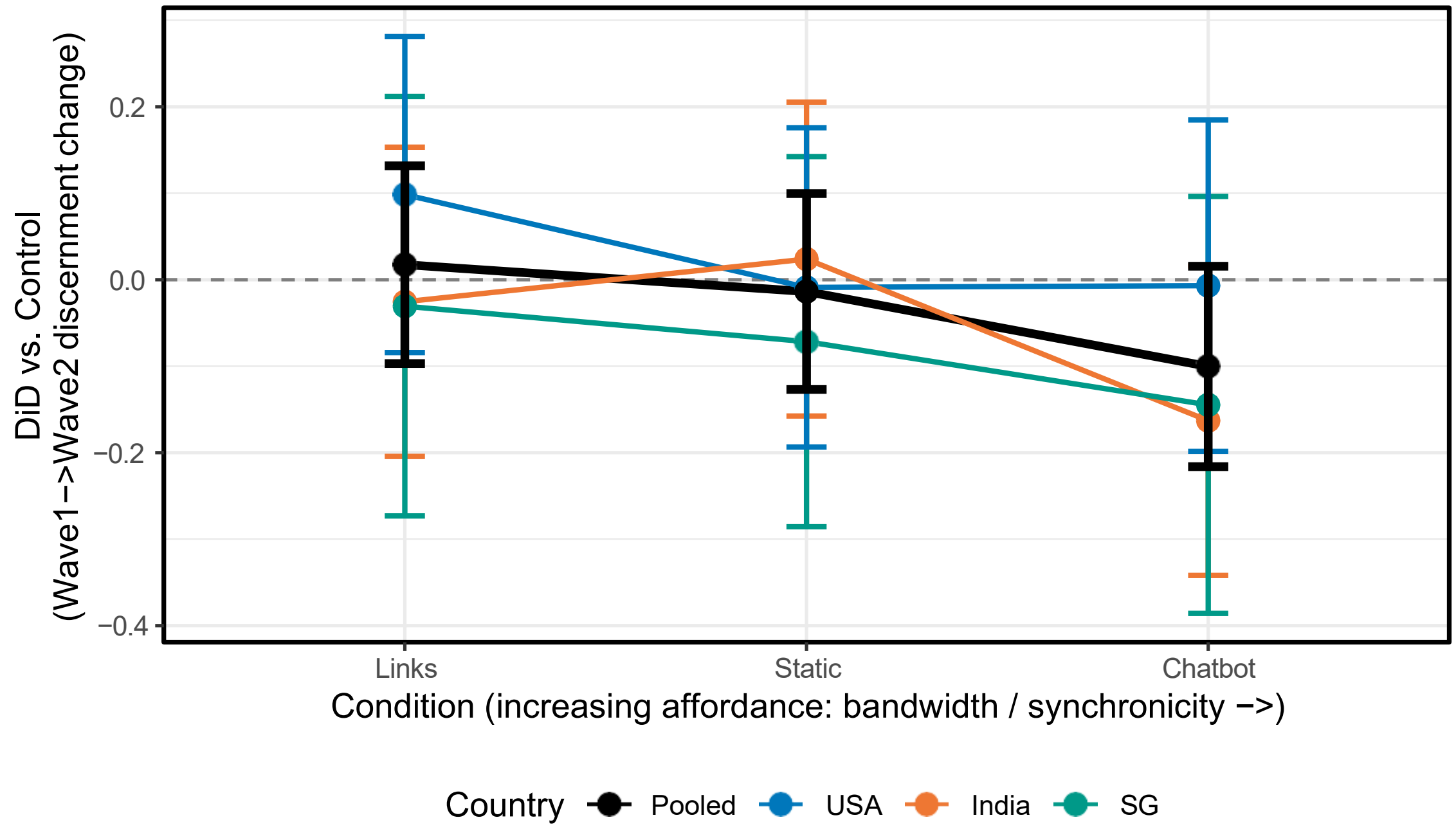

Fig. 9. Wave 1→Wave 2 difference-in-differences in discernment, by country (averaged-item baseline), plotted against affordance level, alongside the pooled estimate (black) already shown in the main text. Negative values indicate the condition's relative advantage decayed.

### F.7 Post-hoc power analysis

Behind the power-analysis figure in the main text's Results.

## G Persistence: Item Definitions

Because Wave 2 assigns a second treatment partway through, our persistence analysis (main text, Results, "Metalearning at One-Week Follow-Up") uses only Wave 2 items answered *before* that second randomization ("block 1").Block 1 item sets, and the resulting linked analysis sample sizes, are:

## H Power analysis

Fig. 10. Post-hoc achieved power for key comparisons, using observed effect sizes and sample sizes (unequal-$n$ two-sample tests). Bars reaching the 80% convention are highlighted. Full-width reproduction of Figure 6.

| India participant (“metro reached the cities” claim) | U.S. participant (“Amish voters for Trump” claim) |
|---|---|
| BOT: Hello, I am a fact-checking assistant. The image we are looking at appears to be from a specific event. First let’s test the text-image consistency. Can you describe what you see in the image, including any visible signs, banners, or notable landmarks that could help identify the location and nature of the event? | BOT: Hello, I am a fact-checking assistant. The image we are looking at appears to be from a specific event. First let’s test the text-image consistency. Can you describe what you see in the image, including any visible signs, banners, or notable landmarks that could help identify the location and nature of the event? |
| PARTICIPANT: I see only metro train and tracks and some palm trees | PARTICIPANT: I believe that was a good topic. |
| BOT: Aha! Gotcha. I am an assistant to help you fact-check this information. Now, let’s look at the context of the image. What are the reasons why you think the image matches the claim **If employment hasn’t increased, how did the metro reach the cities of India? Congress will say, BJP will do!**? | BOT: Aha! Gotcha. I am an assistant to help you fact-check this information. Now, let’s look at the context of the image. What are the reasons why you think the image matches the claim **Over 200,000 Amish voters registered to vote for Trump in Pennsylvania in 2024**? |
| PARTICIPANT: Yeah right | PARTICIPANT: If I recall it help a lot. |
| BOT: Thanks for sharing your thoughts. It seems like you’re skeptical about the claim. Let’s break it down further. Next, let’s examine the **original caption**: Singapore public transport: SBS Transit and SMRT Trains apply for fare increase. Can you explain how this caption relates to the **out-of-context claim**: If employment hasn’t increased, how did the metro reach the cities of India? Congress will say, BJP will do!? What specific parts of the caption or the claim stand out to you, and how do you think they are connected or disconnected? | BOT: Thanks for sharing your thoughts. Next, let’s test the image-evidence consistency. Here is a link to the original news article where this image was published. Here’s an excerpt: **The image shows an Amish community parade in Ohio from 2020, where participants showed support for Trump, not related to voter registration figures in Pennsylvania for 2024**. Now, considering the original caption, image, and excerpt, can you explain how the claim relates to the information provided in the article? |
| PARTICIPANT: It’s essential project so both parties do work. | PARTICIPANT: It provided that they supported Trump and they were doing a March with American flags. |
| BOT: Thanks for sharing! It sounds like you’re considering the potential benefits of a project. Let’s take a step back and examine the original caption again. Remember, we’re trying to fact-check the claim by checking **text-image consistency** and **text-evidence consistency**. What do you think is the main purpose of the original caption, and how does it compare to the claim? | BOT: Thanks for sharing your observations. However, let’s take a closer look. The original caption reads: “Plain Parade” Rallies Support For President In Amish Ohio in 2020. Can you explain how the location and date mentioned in the caption might affect the claim’s validity? Does the claim mention the same state or year as the original caption? |
| PARTICIPANT: Every one has their own perspective, so the government should done | PARTICIPANT: There are a lot of Amish people in the state of Ohio. It makes the March more truthful. |
| BOT: It sounds like you’re highlighting the importance of considering different perspectives. Let’s summarize what we’ve discussed so far: we analyzed the image of a metro train and found that it didn’t provide direct evidence for the claim about employment and metro development; we examined the original caption, which mentioned a fare increase in Singapore’s public transport, and found that it wasn’t related to the claim at all. Would you like to continue exploring this topic, or close the conversation? | BOT: That’s a good observation, but the claim specifically mentions Pennsylvania and 2024, while the original caption and article mention Ohio in 2020. This raises an inconsistency between the claim and the evidence. Can you think of a possible explanation for why the claim might not be entirely accurate – for example, might it be using an unrelated image to support a different statement? This is the last step before I summarize our findings. |
| | PARTICIPANT: There has been so much information givin by many different people. |
| | BOT: Let’s summarize our findings: the original image depicts an Amish community parade in Ohio from 2020, where participants showed support for Trump; the claim’s location (Pennsylvania vs. Ohio) and date (2024 vs. 2020) do not match the original caption; based on our analysis, the claim is likely **misleading or inaccurate** – the image is being taken out of context to support a different statement. |

Table 3. Example conversations with the dynamic fact-checking chatbot. The left example is from an India participant on an out-of-context political claim (a Singapore transit-fare image repurposed as an Indian metro-employment claim); the right example is from a U.S. participant on an out-of-context claim about Amish voter registration (an Ohio 2020 parade photo repurposed as a Pennsylvania 2024 claim). Both illustrate the chatbot’s five-step guided fact-checking process (text-image consistency, text-evidence consistency, source verification, evidence evaluation, and summary).

Table 4. Wave 1 randomization by country and condition, tallied directly from each country’s raw treatment-assignment column.

| Country | Control | Links | Static | Chatbot |
|---|---|---|---|---|
| India | 397 | 390 | 389 | 384 |
| Singapore | 304 | 311 | 321 | 284 |
| USA | 357 | 386 | 390 | 369 |

Table 5. Wave 1 fielding windows by country.

| Country | Wave 1 fielding window |
|---|---|
| USA | Oct 16–30, 2024 |
| India | Jan 15–30, 2025 |
| Singapore | Jan 7 – Feb 20, 2025 |

Table 6. Wave 1 age-bracket distribution by country.

| Country | Age bracket | % |
|---|---|---|
| USA | 18–24 | 3.7% |
| | 25–34 | 9.2% |
| | 35–44 | 16.0% |
| | 45–54 | 15.0% |
| | 55–64 | 20.5% |
| | 65+ | 35.6% |
| India | 18–24 | 25.1% |
| | 25–34 | 40.1% |
| | 35–44 | 22.1% |
| | 45–54 | 7.9% |
| | 55–64 | 3.1% |
| | 65+ | 1.6% |
| Singapore | 25–34 | 20.9% |
| | 35–44 | 23.2% |
| | 45–54 | 24.4% |
| | 55–64 | 21.2% |
| | 65+ | 10.2% |

Table 7. Covariate balance across the four Wave 1 conditions, by country.

| Country | Covariate | Test | Statistic | $p$ |
|---|---|---|---|---|
| USA | Age | $F$ (3, 1498) | 1.10 | .348 |
| | Income | $F$ (3, 1498) | 1.19 | .312 |
| | Cognitive reflection | $F$ (3, 1498) | 0.86 | .461 |
| | Gender | $\chi^2$ (3) | 1.71 | .636 |
| India | Age | $F$ (3, 1340) | 0.62 | .601 |
| | Income | $F$ (3, 1340) | 1.48 | .218 |
| | Cognitive reflection | $F$ (3, 1323) | 1.40 | .241 |
| | Gender | $\chi^2$ (3) | 0.83 | .842 |
| Singapore | Age | $F$ (3, 1179) | 0.51 | .674 |
| | Income | $F$ (3, 1216) | 1.02 | .383 |
| | Cognitive reflection | $F$ (3, 1216) | 0.89 | .448 |
| | Gender | $\chi^2$ (3) | 5.88 | .118 |

Table 8. Chatbot-condition compliance by country.

| Country | Assigned | Logged ≥1 conversation | Compliance |
|---|---|---|---|
| USA | 369 | 362 | 98.1% |
| India | 384 | 377 | 98.2% |
| Singapore | 284 | 283 | 99.6% |

Table 9. Primary analysis: pairwise $t$-tests of mean discernment (1–5, higher = more accurate) for each treatment vs. Control, by country. Discernment is averaged across the items each participant answered (see analysis/01_build_item_level_data.R for scoring detail).

| Country | Treatment | Mean | Control Mean | $t$ | $df$ | $p$ | Sig | Cohen's $d$ |
|---|---|---|---|---|---|---|---|---|
| Pooled (all 3 countries) | Links | 3.219 | 3.206 | 0.45 | 2248 | 0.650 | | 0.019 |
| | Static | 3.234 | 3.206 | 1.01 | 2261 | 0.312 | | 0.042 |
| | Chatbot | 3.268 | 3.206 | 2.27 | 2198 | 0.023 | * | 0.097 |
| USA | Links | 3.402 | 3.439 | -0.85 | 741 | 0.395 | | -0.062 |
| | Static | 3.394 | 3.439 | -1.01 | 743 | 0.314 | | -0.073 |
| | Chatbot | 3.494 | 3.439 | 1.28 | 723 | 0.199 | | 0.095 |
| India | Links | 3.035 | 3.030 | 0.13 | 889 | 0.899 | | 0.009 |
| | Static | 3.063 | 3.030 | 0.77 | 888 | 0.443 | | 0.051 |
| | Chatbot | 3.084 | 3.030 | 1.29 | 886 | 0.198 | | 0.086 |
| SG | Links | 3.256 | 3.191 | 1.32 | 608 | 0.188 | | 0.106 |
| | Static | 3.277 | 3.191 | 1.81 | 623 | 0.070 | . | 0.145 |
| | Chatbot | 3.260 | 3.191 | 1.41 | 575 | 0.160 | | 0.116 |

Table 10. Sharing intention (1–5, higher = more willing to share misinformation – a successful intervention shows a lower mean than Control) for each treatment vs. Control, by country. False-claim items only.

| Country | Treatment | Mean | Control Mean | $t$ | $df$ | $p$ | Sig | Cohen's $d$ |
|---|---|---|---|---|---|---|---|---|
| Pooled (all 3 countries) | Links | 2.543 | 2.698 | -2.84 | 2239 | 0.005 | ** | -0.120 |
| | Static | 2.557 | 2.698 | -2.59 | 2243 | 0.010 | ** | -0.109 |
| | Chatbot | 2.540 | 2.698 | -2.86 | 2198 | 0.004 | ** | -0.122 |
| USA | Links | 2.139 | 2.098 | 0.46 | 731 | 0.643 | | 0.034 |
| | Static | 2.092 | 2.098 | -0.06 | 738 | 0.949 | | -0.005 |
| | Chatbot | 1.971 | 2.098 | -1.44 | 717 | 0.151 | | -0.107 |
| India | Links | 3.151 | 3.346 | -2.35 | 887 | 0.019 | * | -0.157 |
| | Static | 3.184 | 3.346 | -2.00 | 884 | 0.046 | * | -0.134 |
| | Chatbot | 3.173 | 3.346 | -2.13 | 885 | 0.034 | * | -0.143 |
| SG | Links | 2.168 | 2.459 | -3.33 | 611 | <.001 | *** | -0.269 |
| | Static | 2.255 | 2.459 | -2.36 | 617 | 0.019 | * | -0.189 |
| | Chatbot | 2.286 | 2.459 | -1.91 | 584 | 0.057 | . | -0.157 |

Table 11. Confidence in accuracy judgments (1–5, higher = more confident) for each treatment vs. Control, by country. Exploratory outcome; no directional hypothesis assumed. All items (true and false).

| Country | Treatment | Mean | Control Mean | $t$ | $df$ | $p$ | Sig | Cohen's $d$ |
|---|---|---|---|---|---|---|---|---|
| Pooled (all 3 countries) | Links | 3.852 | 3.884 | -0.95 | 2248 | 0.340 | | -0.040 |
| | Static | 3.836 | 3.884 | -1.45 | 2259 | 0.146 | | -0.061 |
| | Chatbot | 3.856 | 3.884 | -0.85 | 2201 | 0.394 | | -0.036 |
| USA | Links | 3.841 | 3.883 | -0.79 | 741 | 0.428 | | -0.058 |
| | Static | 3.842 | 3.883 | -0.81 | 740 | 0.420 | | -0.059 |
| | Chatbot | 3.882 | 3.883 | -0.03 | 724 | 0.975 | | -0.002 |
| India | Links | 4.050 | 4.091 | -0.75 | 888 | 0.452 | | -0.050 |
| | Static | 4.019 | 4.091 | -1.31 | 889 | 0.191 | | -0.088 |
| | Chatbot | 3.978 | 4.091 | -2.09 | 886 | 0.037 | * | -0.140 |
| SG | Links | 3.580 | 3.583 | -0.04 | 607 | 0.969 | | -0.003 |
| | Static | 3.574 | 3.583 | -0.15 | 621 | 0.880 | | -0.012 |
| | Chatbot | 3.631 | 3.583 | 0.83 | 579 | 0.408 | | 0.068 |

Table 12. Secondary analysis: item-level OLS with item fixed effects and participant-clustered standard errors. Reference category is Control. A crossed random-effects model (`lme4`: discernment ∼ treatment + (1|participant) + (1|item)) gives near-identical pooled estimates (see analysis/02_fit_models.R).

| Country | Treatment | Estimate | SE | $t$ | $p$ | Sig |
|---|---|---|---|---|---|---|
| Pooled | Links | 0.004 | 0.026 | 0.16 | 0.874 | |
| | Static | 0.017 | 0.026 | 0.64 | 0.524 | |
| | Chatbot | 0.073 | 0.026 | 2.80 | 0.005 | ** |
| USA | Links | -0.034 | 0.043 | -0.79 | 0.428 | |
| | Static | -0.035 | 0.044 | -0.79 | 0.429 | |
| | Chatbot | 0.060 | 0.043 | 1.41 | 0.160 | |
| India | Links | 0.011 | 0.041 | 0.26 | 0.793 | |
| | Static | 0.036 | 0.042 | 0.87 | 0.384 | |
| | Chatbot | 0.065 | 0.041 | 1.60 | 0.110 | |
| SG | Links | 0.071 | 0.047 | 1.51 | 0.132 | |
| | Static | 0.093 | 0.046 | 2.05 | 0.041 | * |
| | Chatbot | 0.106 | 0.049 | 2.17 | 0.030 | * |

Table 13. Covariate-adjusted regression: misinformation discernment (1–5, higher = more accurate), by country. Reference category is Control. SG omits Political Ideology (not measured in the SG questionnaire – see analysis/08_build_covariates.R).

**USA**

| Variable | Estimate | SE | *t* | *p* | Sig |
|---|---|---|---|---|---|
| (Intercept) | 2.930 | 0.075 | 39.30 | <.001 | *** |
| Links | -0.019 | 0.042 | -0.46 | 0.646 | |
| Static | -0.030 | 0.042 | -0.71 | 0.477 | |
| Chatbot | 0.056 | 0.042 | 1.32 | 0.185 | |
| Cognitive Reflection | 0.353 | 0.039 | 8.99 | <.001 | *** |
| Political Ideology | -0.010 | 0.009 | -1.11 | 0.265 | |
| Gender | 0.027 | 0.030 | 0.91 | 0.364 | |
| Age | 0.105 | 0.010 | 10.49 | <.001 | *** |
| Income | -0.009 | 0.011 | -0.78 | 0.436 | |
| $N = 1496$, $R^2 = 0.142$ | | | | | |

**India**

| Variable | Estimate | SE | *t* | *p* | Sig |
|---|---|---|---|---|---|
| (Intercept) | 3.071 | 0.092 | 33.45 | <.001 | *** |
| Links | 0.045 | 0.049 | 0.92 | 0.360 | |
| Static | 0.066 | 0.050 | 1.32 | 0.188 | |
| Chatbot | 0.092 | 0.051 | 1.81 | 0.070 | . |
| Cognitive Reflection | 0.162 | 0.041 | 3.99 | <.001 | *** |
| Political Ideology | -0.021 | 0.009 | -2.24 | 0.025 | * |
| Gender | -0.045 | 0.036 | -1.27 | 0.204 | |
| Age | 0.004 | 0.017 | 0.26 | 0.798 | |
| Income | 0.022 | 0.016 | 1.40 | 0.162 | |
| $N = 1211$, $R^2 = 0.025$ | | | | | |

**SG (no political control)**

| Variable | Estimate | SE | *t* | *p* | Sig |
|---|---|---|---|---|---|
| (Intercept) | 3.078 | 0.091 | 33.66 | <.001 | *** |
| Links | 0.046 | 0.050 | 0.93 | 0.351 | |
| Static | 0.087 | 0.049 | 1.76 | 0.079 | . |
| Chatbot | 0.066 | 0.051 | 1.30 | 0.193 | |
| Cognitive Reflection | 0.176 | 0.042 | 4.19 | <.001 | *** |
| Gender | 0.048 | 0.036 | 1.36 | 0.175 | |
| Age | 0.005 | 0.014 | 0.33 | 0.744 | |
| Income | -0.001 | 0.007 | -0.18 | 0.858 | |
| $N = 1183$, $R^2 = 0.020$ | | | | | |

Table 14. Covariate-adjusted regression: sharing intention (1–5, higher = more willing to share misinformation), by country. Reference category is Control. SG omits Political Ideology (not measured in the SG questionnaire).

| **USA** | | | | | |
|---|---|---|---|---|---|
| **Variable** | **Estimate** | **SE** | *t* | *p* | **Sig** |
| (Intercept) | 3.252 | 0.133 | 24.47 | <.001 | *** |
| Links | -0.011 | 0.074 | -0.15 | 0.881 | |
| Static | -0.055 | 0.074 | -0.74 | 0.458 | |
| Chatbot | -0.148 | 0.075 | -1.96 | 0.050 | . |
| Cognitive Reflection | -0.817 | 0.070 | -11.67 | <.001 | *** |
| Political Ideology | -0.030 | 0.015 | -1.97 | 0.049 | * |
| Gender | 0.176 | 0.053 | 3.31 | <.001 | *** |
| Age | -0.322 | 0.018 | -18.01 | <.001 | *** |
| Income | 0.082 | 0.020 | 4.08 | <.001 | *** |
| $N = 1496$, $R^2 = 0.287$ | | | | | |
| **India** | | | | | |
| **Variable** | **Estimate** | **SE** | *t* | *p* | **Sig** |
| (Intercept) | 3.266 | 0.173 | 18.90 | <.001 | *** |
| Links | -0.155 | 0.093 | -1.66 | 0.096 | . |
| Static | -0.189 | 0.094 | -2.01 | 0.044 | * |
| Chatbot | -0.110 | 0.096 | -1.15 | 0.251 | |
| Cognitive Reflection | -0.608 | 0.077 | -7.94 | <.001 | *** |
| Political Ideology | 0.082 | 0.018 | 4.60 | <.001 | *** |
| Gender | 0.006 | 0.067 | 0.09 | 0.932 | |
| Age | -0.091 | 0.032 | -2.88 | 0.004 | ** |
| Income | 0.003 | 0.030 | 0.09 | 0.928 | |
| $N = 1211$, $R^2 = 0.080$ | | | | | |
| **SG (no political control)** | | | | | |
| **Variable** | **Estimate** | **SE** | *t* | *p* | **Sig** |
| (Intercept) | 2.816 | 0.154 | 18.31 | <.001 | *** |
| Links | -0.261 | 0.084 | -3.12 | 0.002 | ** |
| Static | -0.205 | 0.083 | -2.47 | 0.014 | * |
| Chatbot | -0.136 | 0.086 | -1.59 | 0.112 | |
| Cognitive Reflection | -0.644 | 0.071 | -9.12 | <.001 | *** |
| Gender | 0.086 | 0.060 | 1.44 | 0.151 | |
| Age | -0.124 | 0.023 | -5.29 | <.001 | *** |
| Income | 0.009 | 0.011 | 0.84 | 0.404 | |
| $N = 1183$, $R^2 = 0.101$ | | | | | |

Table 15. Persistence: difference-in-differences in discernment, each treatment's Wave 1→Wave 2 change relative to Control's own change, by country. "Avg. baseline" uses the mean of all Wave 1 items a participant answered; "Last-item baseline" uses only their final false item at Wave 1.

| Country | Baseline | Links | Static | Chatbot |
|---|---|---|---|---|
| USA | Average of items | +0.098 ($p$ = .29) | −0.009 ($p$ = .92) | −0.007 ($p$ = .94) |
| USA | Last item only | +0.267 ($p$ = .11) | −0.245 ($p$ = .15) | +0.054 ($p$ = .76) |
| India | Average of items | −0.026 ($p$ = .78) | +0.024 ($p$ = .80) | −0.163 ($p$ = .075) |
| India | Last item only | −0.239 ($p$ = .13) | −0.258 ($p$ = .087) | −0.528 (p = .0008) |
| SG | Average of items | −0.031 ($p$ = .80) | −0.072 ($p$ = .51) | −0.145 ($p$ = .24) |

Table 16. Pooled persistence model: diff. discernment ∼ treatment + country.

| Term | Estimate | SE | $t$ | $p$ |
|---|---|---|---|---|
| (Intercept) | −0.076 | 0.050 | −1.52 | .128 |
| Links | +0.017 | 0.058 | 0.30 | .766 |
| Static | −0.014 | 0.058 | −0.24 | .813 |
| Chatbot | −0.100 | 0.059 | −1.70 | .090 |
| Country: India | +0.214 | 0.047 | 4.55 | <.001 |
| Country: SG | +0.400 | 0.053 | 7.49 | <.001 |

Table 17. Post-hoc achieved power for key comparisons, using observed $d$ and $n$ (unequal-$n$ two-sample tests).

| Comparison | $d$ | $n$ (per group) | Achieved power |
|---|---|---|---|
| W1 discernment, Chatbot (pooled) | +0.097 | 1099 / 1104 | 62% |
| W1 discernment, Chatbot (USA) | +0.095 | 369 / 357 | 25% |
| W1 discernment, Chatbot (India) | +0.086 | 446 / 443 | 25% |
| W1 discernment, Chatbot (SG) | +0.116 | 284 / 304 | 29% |
| W1 sharing, Chatbot (pooled) | −0.122 | 1098 / 1103 | 82% |
| Persistence, avg. items (India) | −0.175 | 198 / 193 | 41% |
| Persistence, last item (India) | −0.378 | 162 / 156 | **92%** |
| Persistence, avg. items (SG) | −0.151 | 125 / 147 | 24% |
| Persistence, repeat item (SG) | −0.468 | 51 / 62 | **69%** |

Table 18. Persistence-analysis block-1 item definitions and linked sample sizes.

| Country | Block-1 items | Linked $N$ |
|---|---|---|
| USA | image14, 18, 13, 20 | 800 |
| India | image9, 15, 13, 6 | 726 |
| Singapore | image4, 7, 10, 15 | 520 |